\documentstyle[aps,graphicx]{revtex}

\begin{document}
\title{ Magneto-localization in  Disordered Quantum Wires  }
\author{ Stefan Kettemann, Riccardo Mazzarello}
\address{I. Institut f. Theoretische Physik, Universit\" at Hamburg, Germany}
\maketitle

\begin{abstract}
 The magnetic field dependent localization in a disordered quantum wire 
 is considered nonperturbatively.
 An  increase  of an averaged  localization length with the magnetic field is 
 found,  saturating at twice its value without magnetic field. 
  The crossover behavior is shown to be governed
 both in the weak and strong localization regime 
 by the magnetic diffusion length 
 $L_B$. This function is derived analytically in closed form as a function of 
  the ratio of the mean free path $l$, 
 the wire thickness $W$,    and the magnetic length $l_B$
 for a two-dimensional wire with specular boundary conditions, as well 
 as for a parabolic wire.  
 The applicability of the 
 analytical formulas to resistance measurements in the strong localization
 regime is discussed. 
 A  comparison  with recent experimental 
 results is  included. 
\end{abstract}

\tableofcontents

\section{INTRODUCTION}
\label{sec:introduc}
The 
phase coherent movement of electrons in  a disorder potential 
 can result in strong localization  due to quantum interference
\cite{reviews,ef}.
 As soon as the localization length 
  $L_c$ becomes smaller than  the size of the sample $L$
and  the phase coherence length, $L_{\phi}$,
the resistance increases exponentially.

  The 
 strong localization  due to quantum interference
 is  known to   
  depend
 on the global symmetry of the  disordered electron system~\cite{larkin}. 
 In disordered quantum wires, 
  the localization length is 
\begin{equation} \label{locsym}
L_c = \beta \pi \hbar \nu S D_0,
\end{equation}
 where $\beta =1, 2, 4$,
 corresponding to no magnetic field, finite magnetic field, and 
  strong spin- orbit scattering  or  magnetic impurities,
 respectively. $ \nu (E) $ is the electronic 
 density of states in  the wire. 
 $ D_0 = \tau v_F^2/d $ is the classical diffusion constant 
 of the electrons in the wire,
 with $\tau$ the elastic scattering time, $v_F$ the Fermi velocity, 
 and $ d$  the dimension of classical diffusion. 
  $S$ is the wire crossection. 
  This result was first obtained by calculating the spatial
 decay of the  density correlation function 
  for  wires with diffuse crossections and many transversal 
 channels $N  \gg 1$.
 It can also be  obtained 
  by calculating the 
 transmission probability through thin, few channel   wires
  upto a correction of order $1/N$: $ L_c = (\beta 
 N + 2 - \beta ) l$  
\cite{reviews}, where $l= v_F \tau$ is the mean free path,
 and  $\beta =1, 2, 4$, as defined  above.
 This correction ensures that the localization length is for 
 a single channel, $N=1$, independent of $\beta$, $L_c = 2 l$.
 
 Recently, the doubling of the localization length was observed 
 in sub-micron thin wires of Si $\delta$- doped Ga As 
 structures by Khavin, Gershenson and Bogdanov,
 who found a continously decreasing activation energy when the magnetic field 
 is increased, saturating indeed at one half of its field free value
\cite{khavin}.
 This symmetry dependence of the localization properties of 
 quantum wires allows to  test our present theoretical 
 understanding by detailed comparison with  the experiment. 
  The quantum wires used in the experiment 
 have mean free paths which are smaller than or  comparable to  
  their thickness. Also, in addition to the disorder in the 
 bulk due to the random electrostatic potential 
 of the   donor impurities, there is an unspecified surface roughness
 which may influence the classical  mobility of the wires 
 as well as  its quantum 
 transport  properties. Therefore, a more detailed analysis of the 
 localization length as function of these parameters is 
 called for, in order to be able to compare the theory 
 with the experimental  results  quantitatively. 

 In the next section we review the known weak localization 
 corrections to the conductivity in disordered quantum wires and
 their magneto-sensitivity as function of mobility, wire thickness, and   
 electron density \cite{altaronov,dugaev,aronov,beenakker}. 

 In the third section,  
 the non-perturbative theory of  localization in disordered
 electron systems\cite{ef} is  extended,  in order to 
 allow  the study of  wires with ballistic crossections. 

 In the fourth section,
 the   magnetic phase shifting rate is introduced and 
 identified with a correlation function of the magnetic vector potential, 
 relating  it  to  the 
 coefficient of the time reversal symmetry breaking term in the nonlinear 
 sigma model. 
  This  expression for the magnetic phase shifting rate,
 is calculated anayltically  for arbitrary ratios of the mean free path 
 $l$ and the width of the wire $W$, and  compared with 
 previously derived analytical and numerical results\cite{dugaev,beenakker}
 for a wire with specular boundary scattering. 
 
Next, it is  calculated for a  wire with harmonic confinement 
 which allows to extend the analysis     to stronger magnetic 
 fields,  when the cyclotron radius,  $l_{\rm C}$ is  smaller than  the 
 the wire thickness $W$,  
 but still larger than the elastic mean free path.
 In that regime a new enhancement 
 mechanism for the magnetic phase shifting rate 
 leading to a stronger  magneto-sensitivity, is identified.

 In the fifth section,  
the autocorrelation function 
 of spectral determinants
 (ASD)\cite{haake,us} is considered for a coherent disordered quantum wire, 
 which shows the expected   crossover from 
 Wigner- Dyson statistics\cite{dyson},
 typical for a spectrum of extended states in 
 phase coherent disordered metal systems\cite{ef}, 
to Poisson statistics, corresponding to a spectrum of localized states
\cite{efetov1982,altshk,isa,altfuchs,mirlin,guhr},
  as the length of the wire is increased
 beyond a localization length $L_C$, as reported
  earlier\cite{prb}.   

 This crossover length scale to Possionian statistics is 
 used to derive the  averaged localization length of 
 disordered quantum wires, 
 and it is shown that it  yields the  correct 
  symmetry dependence,  
 Eq. (\ref{locsym}). 
 A comparison with the result of the 
 supersymmetric theory of the two-terminal 
 conductance of a disordered quantum wire, is given. 
 It is concluded, that  the definition of an averaged localization length, 
 by the decay of an energy level correlation 
 function, can be used 
 to consider analytically   the magnetic field dependence of the 
 localization length.
  Thereby,  analytical formulas for the localization length 
    as     
 a  function of wire width, mean free path and magnetic field 
  are derived.

  In the sixth section,
 the theory of finite tempreature magnetoresistance  in
 quantum wires is
 discussed.  In particular,
 the variable range hopping conductivity 
 in quantum wires is reviewed 
 for various temperature and dimensional 
 regimes.  
  It is shown  that in 
 a wide  temperature regime the resistance
  has an activated behaviour, and that therefore, the activation gap
 can be directly measured and related to the localization length 
 of the electrons in the wire.

 This allows a comparison of  the  analytical results for the 
 magnetic field dependence of the 
localization length  with these experimental results, as done in the seventh
 section. 

 In appendix A, the functional integral representation of the ASD
 by Grassmann intergals is given, and the averaging over disorder is
 performed.  
 In appendix B the derivation of the magnetic phase shifting rate is
 given. 
 In appendic C the representation of the matrix fields
 $Q$ is given, and their Laplacian derived. 

\section{  Weak   Localization }

 Classically, the transport of a disordered conductor is characterized 
 by its mobility $\mu = q \tau/m$ and the electron density 
 $n$ related to the  the classical Drude conductivity
 $ \sigma_0 = n q^2 \tau/m$.
 Alternatively, it can be characterized by the diffusion constant 
 $D$,  which is in a metal related to the conductivity by the 
 Einstein relation $ \sigma_0 = 2 q^2 \nu D $.   
 
 When  the electrons diffuse coherently,   
  quantum interference   without magnetic field results in 
 a suppression of the conductivity of a quantum wire of
 order\cite{reviews,ab,gorkov,khm,elk,schmid}

\begin{equation}
\frac{\Delta \sigma}{\sigma_0} = - \frac{2 }{\sqrt{2 \pi^3 }} ( \frac{\sqrt{\tau_{\varphi}}}{\sqrt{\tau}} - 1 ),    
\end{equation} 
where 
 $\tau_{\phi}$ is the phase coherence time, that increases when 
 decreasing  the temperature as a power law:
\begin{equation}
 \tau_{\phi} \sim T^{-\gamma}
\end{equation} 
 and defines the phase coherence length, which an electron 
 diffuses coherently, 
 $ L_{\phi} = ( D   \tau_{\phi} )^{1/2} $. 

 Quasi elastic  electron-electron scattering
 can be the dominant low temperature dephasing mechanism and 
 yields $\gamma = 2/3$ for a 1-d wire and $\gamma=1$ for a 2-d film
\cite{aronov,phase2}. 
At higher temperatures the exponent crosses over to 
 $\gamma = 4$ due to electron-phonon scattering at temperatures
 $ k_{\rm B} T \ll ( \hbar^2/\tau \epsilon_F) \Omega_D$
 where $\Omega_D$ is the optical Debye phonon frequency.
 This power can be smaller,  due to the confinement, in quantum wires. 
 
 The above  definition of the 
 phase coherence rate   is not  applicable 
  when approaching the 
 localized regime, and the  phase coherence length is larger than the 
 localization length $L_c$.
 Also,  there are   mechanisms which may lead to  a saturation 
 of $\tau_{\phi}$  below $T=1 K$, as observed in a wide
 range of conductors \cite{phase1,buettikerphase}.

 A magnetic field breaks the time reversal symmetry.
 Therefore, the magnetic phase accumulated in a  
 Brownian motion of electrons, enters   effectively 
  as an additive contribution to the 
  phase coherence rate, diminishing 
 the weak localization corrections of the conductivity\cite{khm}. 
 For wires with diffusive width $W > l$, it varies quadratically 
 with the magnetic field,  
 $1/\tau_{\phi} (B) = 1/\tau_{\phi} + D \frac{q^2}{\hbar^2} S B^2/K_D$,  
  where S is the crossection of the wire, and the constant 
 $K_D$ depends on the geometry of the wire,
 the direction of the magnetic field  and the scattering mechanisms
\cite{altaronov}. For example, for a 2-dimensional wire of diffusive
 crossection in a perpendicular magnetic field, it yields, $K_D=3$.
 In this way, the conductivity increases
 to its classical value, when the magnetic field is turned on. 

 For a wire with   ballistic crossection  and a magnetic 
 field being perpendicular to its crossection, the
 magnetic field dependence of the weak localization correction 
 to the conductivity is weakened by flux cancellation effects due to 
 boundary scattering\cite{dugaev}.
 If the magnetic field is so small that less than one 
 flux quantum $ \phi_0 = h/e $ is penetrating an area $ W l$,  
 the effective dephasing rate
 $1/\tau_{\phi} (B) $ is  quadratically increasing as for 
 diffusive crossections.
 Its slope was found to be 
      by  at least a  factor $W/l$ smaller, as a consequence of 
 the flux cancellation effect of edge to edge skipping
 orbits\cite{dugaev,beenakker}. 

 When $B W l \gg \phi_0$,  the effective dephasing rate
 $1/\tau_{\phi} (B) $  
 was found by a semiclassical method, 
  to increase  
 only linearly with the magnetic field $B$
 in this regime\cite{dugaev,beenakker}.

 In the presence of magnetic impurities,  
  scattering the electrons with a rate  $1/\tau_S$, 
there is no 
 temperature dependence of 
 the conductivity, if $1/\tau_S \gg 1/\tau_{\phi}$.

 Strong spin-orbit scattering reverses the sign of the quantum correction 
 to the conductivity\cite{hikamiln}. The conductivity is then 
larger than classically expected.  
 This can be observed by increasing an external magnetic field, which 
 destroys time reversal invariance and acts through an effective 
 decoherence time   $1/\tau_{\phi} (B) = 1/\tau_{\phi}$
  as noted above. In the case of moderately strong spin-orbit scattering,
  the conductivity decreases therefore 
 when the magnetic field is turned on\cite{aronov}. 

  At low temperatures, when the dephasing rate $1/\tau_{\phi}$
 becomes smaller than the  typical energy scale of strong localization, 
 the local level spacing $ \Delta_C = 1/( \nu W L_C)$, 
 a perturbation theory in the elastic scattering rate $1/\tau$
 is no longer appropriate, and a nonperturbative 
 treatment of disorder is called for, as the scaling theory 
 of localization 
 does indicate\cite{ab,gorkov}.

\section{ Nonperturbative theory of 
 localization in disordered quantum wires}

 In this section, the nonperturbative theory of disordered 
 noninteracting electrons in quantum wires 
 is derived\cite{weg,elk,ef}.
 Its action, governed by  
  the  long wave length modes corresponding to diffusion,
   the    nonlinear sigma model  is rederived, extending previous 
 derivations, to allow the description of quantum wires with 
 ballistic crossections. 
  
   The Hamiltonian of disordered noninteracting electrons is 
\begin{equation}
H= \epsilon \left({\bf p
}- q {\bf A} \right) + V({\bf x}) +{\bf \sigma  b_s ( x) }
+ {\bf \sigma u_{SO}} \times { \bf p
}  ,
\end{equation} 
 where $q$ is the electron charge.
 In the following, we will generally approximate the
 electronic dispersion   
 $ \epsilon \left({\bf p
}- q {\bf A} \right)$ by  $\left({\bf p
}- q {\bf A} \right)^2/(2 m)$, where $m$ is the effective electron mass, 
 but note that higher moments are sometimes needed to regularize
 the  correlation functions, calculated below. 

 $V({\bf x})$ is taken to be a Gaussian distributed random function
 $\left<V({\bf x})\right> = 0$, 
 and 
 $\left<V({\bf x}) V({\bf x'})\right> =\hbar \Delta S L/(2 \pi \tau) 
 \delta ({\bf x} - {\bf x'}),$
 which models randomly distributed, uncorrelated impurities in the sample.
 $\Delta= 1/(\nu S L) $ is the mean level spacing.  
 This corresponds to a Gaussian distribution function 
 $ P( V ) = \exp ( - \frac{\pi \tau}{ \hbar \Delta  }
 \int \frac{d {\bf x}}{Vol.}
 V ( {\bf x} )^2)$ 
  of the disorder potential,
  defining the  disorder average as 
  $<...>_V = \int \prod_{{\bf x}}  d V P(V) ... $. 
  According to the central limit theorem, this is
 therefore  
 a good description of the various sources of randomness in the 
 electrostatic potential, in which the electrons are moving. 

 The vector potential is used in 
 the gauge ${\bf A} = (- B y, 0, 0 )$, where $x$ is the coordinate
 along  the wire of length L,
  $y$ the one  in the direction perpendicular both 
 to the wire
 and the magnetic field ${\bf B}$,
 which is directed  perpendicular to the wire. 
 The angular brackets denote averaging over impurities.  
 ${\bf \sigma }$ is the electronic spin operator, and 
 ${\bf  b_s ( x) }$ is a random  magnetic impurity  field. 
 ${\bf u_{SO}}$ is the 
local electrostatic field of impurities with large atomic   number $Z$, 
 which do give a stronger spin orbit 
 coupling to the conduction electrons. 

 The Hamiltonian can be classified by its symmetry with respect to 
 time reversal and spin rotation as summarized in Table 1.

 It has been noted that the averaged density of states
 or the averaged one-particle Green's function does not contain 
 any information on the localization of Eigenfunctions of the 
 disordered Hamiltonian $H$\cite{weg}. 
   The physical reason is,  that the one-particle Green's function 
 describes the propagation of the wave function amplitude 
 $\psi ( {\bf x} )$.  Elastic impurity scattering
  randomizes  the phase of the amplitude
 and therefore, this propagator decays  on the scale 
 of the mean free scattering time $\tau$. To catch classical diffusion 
 and quantum localization, at least the 
 evolution of the density or amplitude square has to be averaged 
 over the disorder, leading to a correlation function of two 
 one-particle Green's functions. 
 While  weak localization corrections can be calculated  within 
 a diagrammatic perturbation expansion of such correalation functions
\cite{altaronov,schmid},   the study of  strong electron 
localization in a  disordered potential,
 necessitates  a  
  nonperturbative averaging of such  products of Green's functions.
  This   
can be achieved  by means of the super-symmetry method, whereby the 
 product of Green's functions is written as a functional integral\cite{ef}. 
 Thus, the average over the form of the disorder potential 
can be done  right at the 
 beginning  as a Gaussian integral, exactly. 

 Here, for simplicity, we  present   the derivation of  
 a simpler correlation function, which does not 
 necessitate the use of the 
 full super-symmetry method, but still contains
 some information on  strong quantum localization, as shown recently\cite{prb,prl,prbr}.

 The 
  statistics of  discrete energy levels
 of a finite coherent, disordered metal particle 
 is an efficient way  to characterize its properties~\cite{ef}.
 This can be studied by calculating a disorder averaged 
autocorrelation function   
between two energies at a distance
 $\omega$ in the energy level spectrum.
 Thereby,  an uncorrelated spectrum of localized states 
 can be distinguished from a  correlated  
 spectrum of extended states. 

 The 
 autocorrelation function of spectral 
determinants (ASD) is  
  the  most simple such 
 spectral  correlation function,  
  which allows to explore complex quantum systems 
  analytically, and still does contain nontrivial information 
 on level statistics and,  thus,  on 
 localization\cite{prb,prl}.  

It is an oscillatory function whose amplitude decays
with a power law, when the energy levels in the vicinity of 
 the central  energy $E$ are extended, while  a Gaussian 
 decay is a strong indication that all states 
are localized.

 It is defined by   
$
C(\omega) = \bar{C}(\omega)/ \bar{C}(0),
\hspace{.5cm} 
\bar{C}(\omega) = \left<\mbox{det}( E + \omega/2 - H) \mbox{det}
 ( E- \omega/2 -H )\right>,
$ where  $E$ is  a central energy. 

      Since it is  a product of two spectral determinants,
 and a spectral determinant can be written as a Gaussian functional integral
 over Grassmann variables $\psi$, $\psi^*$, 
  one does need at least a 2-component Grassman field, 
 one for each spectral determinant. 
 
In general,  
 $4 \alpha$ -component Grassman fields are needed to 
 get the functional integral representation of the ASD.
 Here,  $\alpha=1$, when the Hamiltonian is independent of the spin 
 of the electrons, and each level is doubly spin degenerate. 
  There is  one pair of Grassman fields for each determinant 
 in the ASD and each pair is composed of a Grassman field
 and its time reversed one, as obtained by complex conjugation.
 $\alpha=2$ has to be considered, when the Hamiltonian does depend
 on spin, as for the case with moderately strong magnetic impurity or
 spin- orbit scattering. 
 This necessitates the use of a vector of  a spinor and the corresponding
time reversed one.


  The representation  as a Gaussian functional integral
 over Grassmann variables is given explicitly for $\alpha=1$ in appendix A. 
 There, the averaging over disorder and the decoupling 
 of the resulting  $\psi^4$ interaction with  a Gaussian integral 
 over a matrix field $Q$ is given.
   Thus, the disorder averaged ASD is given 
  by a functional integral over a matrix field $Q$.  

 The matrix $Q$ is element of the full symmetric 
 space, including rotations between the 
 subspace corresponding to the 
left and the right spectral determinant. 
Therefore,   the long wavelength modes
 of $Q$, do contain the nonperturbative   information on the 
diffuson and Cooperon modes. 

 In order to consider the action of  long wavelength 
 modes governing the physics of diffusion and localization, 
 one  can now expand around the saddle point solution of the action,
 satisfying  for $\omega =0$, 
 
\begin{equation} \label{saddle}
 Q =  i/(\pi \nu) < {\bf x} \mid 
 1/( E - H_0 + i \hbar/(2 \tau) Q )
\mid {\bf x} >. 
\end{equation}  
 This saddle point equation is found to be 
 solved by $ Q_0 = \Lambda$. 
For $\alpha=1$, and $B=0$, at $\omega =0$, the rotations $ U$, 
 which leave $Q$ in the  symplectic  symmetric 
 space  yield the complete manifold 
 of saddle point solutions as
 $ Q = \bar{U} \Lambda U$, where $ U \bar{U} = 1$, 
 with  $Q^T C = C Q$. 
 The modes which leave $\Lambda$ invariant,
  elements of $Sp(1) \times Sp(1)$
 are surplus,
 or spontanously broken,  and 
 can be factorized out, leaving the 
 saddle point solutions to be elements of the 
  symmetric space 
 $ Sp(2)/( Sp(1) \times Sp (1)) $\cite{lie}. 

 For $\alpha=2$    the matrix $C$ is, due to the time reversal 
 of the spinor,  substituted 
 by $i \sigma_2 \tau_1$\cite{elk}.
 Both magnetic impurities and 
 spin-orbit scattering   reduce the Q matrix to unity in spin space.
 Thus, C has effectively the form $\tau_1$.
   The condition $Q^T C = C Q$ leads therefore to a new symmetry class, 
 when the spin symmetry is broken
 but the 
 time reversal symmetry remains intact.
 This is the case 
 for moderately strong spin-orbit scattering. Then, $Q$ are $4 \times 4$-
 matrices on the orthogonal 
 symmetric space $ O(4)/(O(2) \times O(2) )$
 \cite{weg}, which is the nonperturbative consequence of the 
 sign change  of a spinor component  under time reversal operation, 
 which leads to the positive quantum correction
 to the conductivity  in perturbation theory
\cite{schmid}. 
  With magnetic impurities both the spin and  time reversal
 symmetry is broken,  and the Q- matrices
 are in the unitary symmetric space $U(2)/(U(1) \times U(1) )$
 as  for a moderate magnetic field and spin degenerate
 levels. The difference in the prefactor 
 $\alpha$  remains.
 One can extend this approach to other compact
 symmetric spaces with physical realizations, 
 see Ref. \onlinecite{zirnbauer,tbfm} for a complete classification.

 In addition to these gapless transversal modes there 
 are massive longitudinal modes with $Q^2 \neq 1$, which  for 
 $N  \gg 1$, can be integrated out\cite{ef}, and  
 the ASD thereby  reduces to  a functional integral over the  transverse
 modes $U$.
 Now, the action of  finite frequency $\omega$ and spatial fluctuations
 of $Q$ around the saddle point solution can be found 
 by an expansion of the action $F$, Eq. (\ref{exactfree}). 
 Inserting $ Q = \bar{U} \Lambda U$  into Eq. (\ref{exactfree}), 
 and performing the cyclic permutation 
 of $U$ under the trace $Tr$, yields, 
 \begin{equation}
 F = - \frac{1}{2} \int  d {\bf x}
 <{\bf x} \mid Tr \ln  (
 G_0^{-1} - U [ H_0, \bar{U} ] + \omega  U \Lambda  \bar{U} ) \mid {\bf x} >,
 \end{equation} 
 where 
\begin{equation}
G_0^{-1} = E - H_0 + \frac{i \hbar }{2 \tau} \Lambda. 
\end{equation}

Expansion to first order in the energy difference 
$\omega$                                      
and to second order in the  commutator 
$U [ H_0, \bar{U} ]$, 
 yields, 
\begin{eqnarray}\label{free2}
F[U]  &=&
- \frac{1}{2}  \omega \int  d {\bf x} <{\bf x} \mid
 Tr G_{0 E} U \Lambda \bar{U} \mid {\bf x} >
 \nonumber \\ &+&
\frac{1}{2}\int  d {\bf x} <{\bf x} \mid Tr G_{0 E} U [ H_0, \bar{U} ] \mid {\bf x} >
\nonumber \\ &+&
\frac{1}{4}\int  d {\bf x} <{\bf x} \mid Tr (G_{0 E} U [ H_0, \bar{U} ])^2 \mid {\bf x} >.
\end{eqnarray}

Note that 
$ [ H_0, \bar{U} ] = - \frac{\hbar^2}{2 m} ({\bf \nabla}^2 \bar{U} )
- \frac{\hbar^2}{ m}({\bf \nabla} \bar{U} )
{\bf \nabla} - \frac{q \hbar }{ i m c }
 ( \tau_3 {\bf A} {\bf \nabla} \bar{U} -  \bar{U} \tau_3 {\bf A} {\bf \nabla})
$. 

 The first order term in 
$U [ H_0, \bar{U} ]$ vanishes 
 for Gaussian white noise isotropic scattering. 

  In general, in order to account for the ballistic motion 
 of electrons in ballistic wires, or to account for different sources
 of randomness, a directional dependence of the matric $ U = U ( {\bf x}. 
{\bf n} ) $, where $  {\bf n} = {\bf p }/\mid {\bf p} \mid$, 
 has to be considered\cite{taras1,blanter}. 
  However, for the geometries considered in this article, we have 
 found that
 the form of the action derived below 
 remains valid for diffusive as well as ballistic crossections, 
 when  the vector fields ${\bf S}$ 
  as intorduced in 
  Refs. \onlinecite{taras1,blanter}, are integrated out.  This 
 will be presented in more detail 
 in a separate article.  

 Then, one can  keep 
 second order terms in ${\bf \nabla} \bar{U}$ and $ {\bf A} $, 
 which turns out to be valid for the   regime of weak disorder,  
 $ l \gg 1/k_F$ and for any magnetic field,  $l_B \gg k_F$.
 Thus,  
 one gets, using the saddle point equation,  Eq. (\ref{saddle}), 
\begin{eqnarray}\label{free3}
F[U]  &=&
- \frac{\pi  }{4} \frac{  \omega}{\Delta } \int \frac{ d {\bf x}}{S L}
 Tr   \Lambda Q  
\nonumber \\ &+&
\frac{1}{4}\int  d {\bf x}  <{\bf x} \mid  Tr 
(G_{0 E} U (  \frac{\hbar^2}{2 m}({\bf \nabla} \bar{U} )
({\bf \nabla} -\frac{i}{\hbar} q {\bf A} \tau_3) + \frac{q \hbar }{  m  }
 [\tau_3, \bar{U} {\bf A} {\bf \nabla}]) )^2 \mid {\bf x} >.
\end{eqnarray}

 Next, one can separate the physics on different length scales, 
 noting that the physics of diffusion and localization 
 is governed by spatial variations
 of $U$ on length scales larger than the mean free path 
 $l$. The smaller length scale physics, is then included in the 
 correlation function of Green's functions, being related to the 
 conductivity by the  Kubo-Greenwood formula, 
\begin{equation}\label{kubo}
\sigma_{\alpha \beta } ( \omega ) = \frac{\hbar}{\pi S L}
\frac{q^2}{m^2}\sum_{\bf p, p'}  (p_{\alpha} - q A_{\alpha})
( p'_{\beta} - q A_{\beta} )
< {\bf p} \mid G^R_{0 E} \mid {\bf p'} > 
< {\bf p'} \mid G^A_{0 E+\omega} \mid {\bf p} > , 
\end{equation}
 where ${\bf p} =\frac{\hbar}{i} {\bf \nabla}$. 
 The remaining averaged correlators, involve products 
 $ G^R_{0 E} G^R_{0 E+\omega}$ and  $ G^A_{0 E} G^A_{0 E+\omega}$
 and are therefore by a factor $\hbar/(\tau E)$ smaller than the 
 conductivity, and can be disregarded for small disorder 
 $\hbar/\tau \ll  E $. 
 In the bulk of this 
  article we are interested in the weak magnetic field 
 limit, where $ \omega_c \tau \ll 1$, with the 
 cyclotron frequency $  \omega_c = q B/m$. 
  In this  limit we can 
 disregard the nondiagonal Hall conductivity and  the explicit 
 magnetic field dependence 
 of the longitudinal conductivity. 

In order to insert the Kubo-Greenwood formula in  the 
 saddle point expansion of the nonlinear sigma model, 
 it is convenient to rewrite the propagator in 
$F$ as
$ G_{0 E} = \frac{1}{2} G^R_{0 E} ( 1 + \Lambda ) + 
\frac{1}{2} G^A_{0 E} ( 1 -  \Lambda) $. 

 Then, we can use,  that  
$Tr [ \sum_{\alpha = 1}^d \sum_{s = \pm} ( 1 + s \Lambda ) U ( \nabla_{\alpha}  \bar{U})
( 1 - s \Lambda ) U ( \nabla_{\alpha}  \bar{U})]
 = - Tr[ ( {\bf \nabla} Q)^2]$, 
 and 
$ Tr [ \sum_{s= \pm} ( 1 + s \Lambda ) U   [\tau_3, \bar{U}]]
( 1 - s \Lambda ) U [\tau_3, \bar{U}]
 = - Tr [ [\tau_3, Q ]^2 ] $. 
 
Thereby we can rewrite Eq. (\ref{free3}) as

 \begin{eqnarray}\label{free4}
F[Q]  &=&
- \frac{\pi  }{4} \frac{  \omega}{\Delta } \int \frac{ d {\bf x}}{S L}
 Tr   \Lambda Q  
\nonumber \\ &-&
\frac{1}{4}\int  d {\bf x}  Tr[ ( {\bf \nabla} Q({\bf x} ))^2
 <{\bf x} \mid  
G_{0 E}^R  \frac{\hbar^2}{2 m} 
({\bf \nabla} - \frac{i}{\hbar} q {\bf A} )
G_{0 E}^A  \frac{\hbar^2}{2 m} 
({\bf \nabla}  - \frac{i}{\hbar} q {\bf A} )
 \mid {\bf x} >
\nonumber \\ &-&
 \frac{1}{4} (\frac{q \hbar }{  m  })^2 
\int  d {\bf x}  Tr[ [\tau_3, Q({\bf x}) ]^2 ]
 <{\bf x} \mid  
G_{0 E}^R   {\bf A}
{\bf \nabla} 
G_{0 E}^A    {\bf A}
{\bf \nabla} 
 \mid {\bf x} > + c.c.
.
\end{eqnarray}

 For wires of thickness $ W $ not exceeding 
 the length scale $L_{C U} = L_C (\beta = 2) = 2 \pi \hbar \nu S D_0$, 
  the variations of the field $Q$ 
 can be neglected in the transverse direction, and the 
 action reduces to the one of a one- dimensional nonlinear sigma model.
 Using the Kubo formula, Eq. (\ref{kubo}),  
  this functional of $Q$ thus simplifies, for $\omega_c \tau \ll 1$, to, 
\begin{eqnarray} \label{freeh}
F = \frac{\pi  \hbar}{16 q^2} \sigma (\omega=0) W
\int_0^L d { x} ( Tr ( { \nabla_x} Q({ x} ))^2
- < A_x \bullet A_x >
  \frac{q^2}{\hbar^2}  Tr [\tau_3, Q({ x}) ]^2 ). 
\end{eqnarray} 
 The prefactor
of the  time reversal symmetry breaking term, the correlation function 
\begin{eqnarray} 
 < A_x \bullet A_x > &=& B^2 < y \bullet y > \nonumber \\
& = & 
\frac{(<{\bf x} \mid  
G_{0 E}^R   {\bf A}
{\bf \nabla} 
G_{0 E}^A    {\bf A}
{\bf \nabla} 
 \mid {\bf x} > + c.c.)}
{<{\bf x} \mid  
G_{0 E}^R   
({\bf \nabla}  - \frac{i}{\hbar} q {\bf A})
G_{0 E}^A   
({\bf \nabla}  - \frac{i}{\hbar} q {\bf A})
 \mid {\bf x} >},
\end{eqnarray}
 is 
 increasing with the magnetic field $B$, suppressing  
 modes with $ [ Q, \tau_3 ] \neq 0 $, the Cooperon modes, 
 arising from the self interference of closed diffusion paths. 
 Accordingly, the symmetry of the $Q$- fields is broken 
 from $Sp(2)/(Sp(1) \times Sp (1) )$ to $U(2)/( U(1) \times U(1) )$. 

 In the next section it is shown that this prefactor is related 
 to the magnetic phase shifting rate, and is evaluated for a 
 disordered quantum wire. 

\section{The Magnetic Phase Shifting Rate}

It can be seen that the  prefactor
 of the symmetry breaking term in  Eq. (\ref{freeh})   is 
 proportional to the effective phase shifting rate $1/\tau_B$, governing  
 the weak localization suppression by a magnetic field. 
 To this end,  one can use the supersymmetric version 
 of the above nonlinear sigma model,
 obtained by substituting the matrix $Q$ by supermatrices, 
 and the trace over matrices $Tr$ by the 
 supertrace $STr$, but keeping all coefficients the same 
 as in Eq. (\ref{freeh}). 
 Then,  the 
weak localization  corrections to the 
 conductivity can be calculated 
as outlined in \cite{ef},
 by an expansion of $Q$ around the classical saddle point $ Q_c = \Lambda$. 
  Thus, 
  the magnetic  phase shifting rate $1/\tau_B$
 can be identified as, 
\begin{equation} \label{magneticphaseshifting}
1/\tau_B =  4  D \frac{q^2}{\hbar^2}
  < A_x \bullet A_x >,
\end{equation}
where   the Einstein relation $\sigma = 2 q^2 \nu D$ 
of the classical conductivity 
 $\sigma$ 
  to the classical  diffusion constant $D$ has been used. 

\subsection{2D wire with specular boundary conditions}
  The general expression for the correlation function 
   $ < y \bullet y > $, is found
 by inserting the  momentum eigenstates of the wire and 
  summing the correlation functions
 of Green's functions for $l_B \gg W$ in Eq. (\ref{h}). 
 It  
 is thus  obtained to be given for a two dimensional 
 wire of width $W$ in momentum representation  by,  
\begin{equation} \label{h}
 < y \bullet y >  =  \sum_{k_x,k_y,k_y'} 
k_x^2 ( G_{0 E}^R( k_x, k_y )   
G_{0 E}^A ( k_x, k_y' ) + c.c. ) \mid < k_y \mid y \mid k_y' > \mid^2/
  \sum_{k_x,k_y} 
(k_x-\frac{q}{\hbar} A_x  )^2 G_{0 E}^R( k_x, k_y )   
G_{0 E}^A ( k_x, k_y ). 
\end{equation}
 Here, $G_{0 E}^{R/A} ( k_x, k_y ) = ( E - \hbar^2 
( k_x^2 + k_y^2 )/(2 m)  \pm i/(2 \tau) $.  

  Keeping all corrections for finite number of transverse
 channels $N = k_F W/\pi$ and effective mean free path
 $\lambda = k_F l$, in  
 the weak disorder limit $ E \gg \hbar/\tau$, we get  for $ N \gg 1$
 the expression :  
\begin{eqnarray}\label{hexact}
  < y \bullet y > &=& W^2
 ( \frac{1}{12}  K -  \frac{1}{ 2 \pi^2} K_1 
- \frac{\lambda^2}{ \pi^2 N^2} K_2 
\nonumber \\ 
&+&\frac{4}{\pi^4} \frac{\lambda^3}{ N^4}
 \sum_{s =1}^{N}  \frac{s^2}{N^2}
 \sqrt{1- \frac{s^2}{N^2} } Im \sqrt{ \frac{s^2}{N^2} 
+ i \frac{2}{\lambda} }
\tan ( \frac{\pi N}{2} ( \sqrt{ \frac{s^2}{N^2}
 + i  \frac{2}{\lambda} } -  \frac{s}{N} ) ))/K_0 , 
\end{eqnarray}
 where the definition of the  constants $K_i$ is given  in Appendix B .

  Its dependence on the mean free path parameter 
 $\lambda = k_F l$ is shown in Fig. 1. 

\begin{figure} \label{fig1}
\includegraphics[width=0.44 \textwidth]{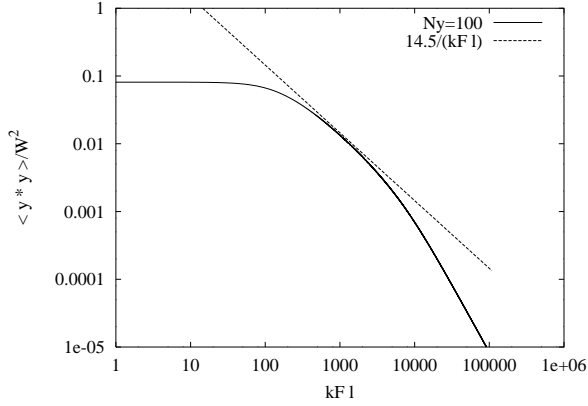}
\begin{center}
\caption{  The dependence of the correlation function 
 $ < y \bullet y >/W^2$
 on the dimensionless 
 mean free path $\lambda = k_{\rm F} l$ for $N=100$ channels. 
  For comparison, the line corresponding to a disorder independent 
 phase shifting rate, approximately valid 
 for $ N \ll  \lambda \ll N^2$, is shown.    }
\end{center}
\end{figure}

 Note that, although
 $N \gg 1$ is required for  the validity of the 
nonlinear sigma model,
 the equation (\ref{hexact})  is valid for arbitrary ratios of the width 
 of the wire  $W$
 and the mean free path $l$, 
 since the motion remains  diffusive along the wire axis on large 
 length scales, even if $l \gg W$.

 For diffusive wire crossections, $l<W$,
$< y \bullet y > \rightarrow \overline{ y^2 } = W^2/12 $ 
 which results 
 exactly in  the  
known   result for the magnetic  phase shifting rate
 $1/\tau_B = 4  D \frac{q^2}{\hbar^2} \overline{y^2}   B^2$
\cite{aronov,beenakker}.

 The above 
 derivation is more general, and applies for arbitary ratios 
 of the  wire  thickness $W$ and  the
 mean free path $l$, as long as 
  the magnetic length $l_B$ is both larger than the width 
 $W$ and the elastic mean free path $l$, and for a large number 
 of transverse channels $ N = k_F W/\pi \gg 1$. 
 
 For ballistic wire crossections, $l>W$,   Eq. (\ref{hexact})
 shows, that the effect of the magnetic field becomes weaker, 
 as  $W/l$ decreases.
   This is a result of the flux cancellation effect, discussed 
 in the limit of weak localization  in Ref. \onlinecite{dugaev,beenakker}: 
 The matrix element of
 the vector potential  $<{\bf k} \mid {\bf A} \mid  {\bf k'}>$
 vanishes for ${\bf k} = {\bf k'}$, since ${\bf A} = (-B y,0,0)$ 
 is antisymmetric in the coordinate perpendicular to the wire, $y$. 
  Thus, elastic impurity scattering is needed to
  mix different momentum states 
 and contribute   finite matrix elements of 
 the magnetic vector potential.

 One can check that   Eq. (\ref{hexact}) is valid also in the weak disorder
 limit, 
 by Taylor expanding   the correlation function in $ 1/(k_{\rm F} l)$,
 giving  
 $< y \bullet  y>= \frac{ W^2}{10} ( N^3/\lambda^2 ) $, showing that it 
 vanishes for $ \lambda \gg N^2$, corresponding to 
 $\hbar/\tau \ll \pi^2 \hbar^2/(2 m W^2)$, when the disorder does not mix 
 transversal modes, like $1/\lambda^2$, as seen in Fig. 1.

 In the intermediate regime, $N < \lambda$,
  it had been argued in
     Ref. \onlinecite{dugaev,beenakker}, that $1/\tau_B$ should be reduced 
 by a factor linear in $N/\lambda$  resulting for a 2 dimensional wire 
 with perpendicular magnetic field in a disorder independent expression 
\begin{equation} \label{ballphase}
\frac{1}{\tau_B} = \frac{1}{C} \frac{W^3 v_F}{l_B^4},
\end{equation}  
 where $l_B = ( \hbar/(q B))^{1/2} $ is the magnetic length. 
  For specular boundary condition, as considered in this article,
  it was found
  numerically that $C= 9.5$\cite{beenakker}. 
 Correspondingly, the function $ < y \bullet  y>/W^2  $ should 
 approach 
 $  < y \bullet  y >/W^2 \rightarrow (\pi/2  C) N/\lambda$ or for $N = 100$, 
  $  < y \bullet  y >/W^2   \rightarrow 16.5/\lambda$. 
 The result Eq. ( \ref{magneticphaseshifting} ) 
 agrees indeed  with this behaviour, in a regime
   $N \ll \lambda \ll N^2 $, 
 although the best fit gives  a different prefactor $14.5$, corresponding 
 to $C = 10.8$.  
  The analytical result  shows,  furthermore, that this behaviour is only 
 an approximation and that there is  a crossover to 
 the perturbative regime, discussed above, where
 $
< y \bullet  y>/W^2
$  decays like  $\sim 1/\lambda^2$,  see Fig. 1. 
 Note that this result is accurate upto corrections of order $1/N$. 
                                                     
\subsection{Parabolic Wire}

 As long as 
 the elastic scattering rate exceeds the cyclotron frequency, 
 $ 1/\tau \gg \omega_c$, or correspondingly, 
 $l \ll   l_{Cyc} $,  where
 $  l_{Cyc} = k_F l_B^2$ is the cyclotron path, determining the length scale 
 on which ballistic paths start to bend due to the Lorentz force, 
  the magnetic field dependence 
 of the  classical diffusion constant and the  density of states 
 can  be neglected, being for a 2- dimensional wire 
 $ D= \tau v_F^2/2$ and $ \nu( E) = m/( 2 \pi \hbar^2)$, respectively. 

 However, the  cyclotron length  can 
  be small compared to the width of the 
 wire, $l_{Cyc}< W$, while exceeding the elastic mean 
 free path  $   l_{Cyc} > l $, when the crossection of the 
 wire is diffusive, $  l < W $ . 
 Thus,   the localization length can depend sensitively on the ratio of 
 these length scales, even in the weak magnetic field limit, 
where the density of states and classical conductivity are insensitive to the 
 magnetic field. 
 In order to study the crossover as function of the magnetic field, 
  the dependence of the eigen functions on the magnetic field have 
 to be taken into account, therefore.
 This regime  is  most conveniently studied 
 for a parabolic wire, having a harmonic confinement, 
 \begin{equation}
 H_0 =\frac{1}{2 m} ( {\bf p} - q {\bf A} )^2 +\frac{1}{2} 
 m \omega_0^2 y^2,
\end{equation}
 having the  energy eigen values 
\begin{equation} 
 E_{n,k} = \hbar \omega_{\rm eff} ( n+ 1/2) + \frac{1}{2 m^*} \hbar^2 k^2,
\end{equation}
 where the effective mass is 
 $ m^* = m \omega_{\rm eff}^2/\omega_0^2$, 
 and the effective frequency is $ \omega_{\rm eff} = 
 ( \omega_{B}^2 + \omega_0^2)^{1/2} $, 
 where $ \omega_{B} = q B/m $ is the cyclotron frequency. 
 The spatial center of the electron eigenstates are shifted
by the guding center $ y_k = k \hbar \omega_B/(m \omega_{\rm eff}^2)$. 
  Thus, the width of the wire is at constant Fermi energy 
$E_{\rm F}$ 
 dependent on the magnetic field $B$. 
   Defining the width of the wire $W$  at fixed Fermi energy as
 $ W^2 = max ( < n,k  \mid y^2 \mid n,k > ) $ with 
 $ E_{n,k} = E_{\rm F}$, one finds for the parabolic wire: 
\begin{equation} 
W^2 ( B) =
l_{\rm eff}^2
{\rm  max} ( 2  \frac{E_{\rm F}}{\hbar \omega_{\rm eff}  }
  \frac{  \omega_{\rm B }^2}{\omega_{\rm 0}^2}
+  (  n +1/2 ) ( 1- 
\frac{  \omega_{\rm B }^2}{\omega_{\rm 0}^2})). 
\end{equation}
  For  large 
 magnetic field,  $ \omega_B \gg \omega_0$, this approaches 
 exactly twice the value at zero magnetic field,  and thus, 
\begin{equation}
 W ( \omega_C \gg \omega_0 )  = \sqrt{2}  W( 0 ) = 
 (2  E_{\rm F}/ ( \hbar \omega_0  ))^{1/2} l_0.
\end{equation} 
 Thus,  the wire width is a slowly vaying function of the paramter
 $ \omega_c/\omega_0 = W( B=0)/l_{cyc}$. 

 The presence of impurities smoothens this function further,  and 
  we can  thus assume the width to be practically magnetic field
 independent: 
\begin{equation}
W =  \sqrt{2  E_{\rm F}/ m_e }/\omega_0.
\end{equation}
  This allows us to study the various regimes of interest as a function of 
 the wire width $W$, the magnetic length $l_{\rm B}$ and  
 the average mean free path 
 $ l = ( 2 E/m )^{1/2} \tau $. 
 
  Naturally, the classical conductivity in such a wire is 
 anisotropic. We find 
 that 
\begin{equation}
\sigma_{xx} = \frac{  1 + \omega_0^2 \tau^2 }{  1 + 
\omega_{\rm eff}^2 \tau^2 } q^2 \tau n_e/ m,
\end{equation}   
and 
 \begin{equation}
\sigma_{yy} = \frac{  1  }{  1 + 
\omega_{\rm eff}^2 \tau^2 } q^2 \tau n_e/ m,
\end{equation} 
where $ n_e = (2/3 \pi ) ( m_e E/\hbar^2 \omega_0) $
 is the average electron  density in  the wire, 
 which is taken  to be  
approximately  independent of the magnetic field. 
 Since 
 we consider magnetic fields where $ \omega_{\rm C} \tau \ll 1$, 
 the classical conductivity is magnetic field independent, 
 $ \sigma_{xx} = q^2   \tau n_e /m$, and
 $ \sigma_{yy} =    \sigma_{xx}/(1 + \omega_0^2 \tau^2) $.

 Thus, the 
condition that the localization is governed by the 
 one-dimensional nonlinear sigma model is changed 
 to $ L_{C U}/( 1 + \omega_0^2 \tau^2 ) > W$. 
 With $ \omega_0 \tau = l/ W$ follows that 
 the one dimensional localization condition requires, $  l < 2 N W$,  
 in the weak disorder regime,  $ k_{\rm F} l \gg 1$.  
 
 Rederiving the nonlinear sigm  model in the representation of a 
 clean parabolic wire,  
  using the definition of the correlation fucntion, Eq. (\ref{h}), 
where    teh sum over transverse momenta is substituted 
 by the sum over the band index, $n$,   $ k_y \rightarrow n$, 
 we find the result, 
\begin{equation} \label{parabol}
 < y \bullet y > = W^2 (\frac{2}{5}(  \frac{1}{ 1 + \omega_0^2 \tau^2 }  +
 3  \frac{\omega_c^2}{\omega_0^2} )
= W^2  \frac{2}{5} (\frac{1}{ 1 + l^2/ W^2  }  +
 3  \frac{W^2}{l_{\rm cyc}^2} ). 
\end{equation}

 Note that, since 
$ \omega_0^2 \tau^2 = l^2/( W^2)$, 
 the ballistic crossection limit $l > W$, coincides 
 for the parabolic wire with the 
 clean wire limit, where transversal modes are not mixed
 by the disorder $ \hbar/\tau < \hbar \omega_0$. 
 Thus the flux cancellation effect
 leads in the parabolic wire to a  supppression of   the 
 phase shifting rate by a factor   $ W^2/l^2$ as found 
 for the wire with specular boundaries in the clean wire limit 
 as seen in the previous subsection. 
 
 Thus, it is not surprising that the behaviour of the 
 magnetic phase shifting rate, as known from 
  weak localization corrections  
  for a wire with ballistic 
 crossection, $ W > l $,  and hard wall boundary conditions, 
 is not reproduced when considering a parabolic wire.  
  In the former case, there is a regime, 
 $W^2 < l_B^2 < W l $, implying 
 $ l_B < l $,  where the magnetic phase shifting         rate is 
 given by 
\begin{equation}
\frac{1}{\tau_B} = \frac{W^2}{ C_2 \tau l_B^2} < \frac{W^3 v_F }{C l_B^4}, 
\end{equation}
 where $C_2 = 24/5$.  This  
 is  smaller than expected from Eq. (\ref{ballphase}), and is not obtained 
 for the parabolic wire.

 Instead,  we find that there is a regime, 
 where   the magnetic field sensitivity of localization 
 becomes   stronger, when the cyclotron length $l_{\rm cyc}$, 
 becomes comparable to the  width 
 of the wire $W$.  When $ l < l_{\rm cyc} < W$
the magnetic phase shifting rate is found to 
 increase with the magnetic field like $B^4$, 
\begin{equation}
\frac{1}{\tau_B} = \frac{24}{5}  D \frac{q^2}{\hbar^2} B^2  
  \frac{W^4}{l_{\rm cyc}^2}. 
\end{equation}

 When the magnetic field becomes so strong that 
 the cyclotron length $l_{\rm cyc}$,  becomes comparable or smaller 
 than the mean free path $l$, or  
 $\omega_c \tau > 1$,  the diffusion constant and the density of states 
 become functions of the magnetic field. Then, 
 the spatial modes of the nonlinear sigma model perpendicular to the 
 wire can 
  become soft and contribute to the functional integral, and 
 thus, the nonlinear sigma model 
  becomes effectively two dimensional. 

 In 
 this limit, a  quantum Hall wire, 
 the approach used in this article  
 can yield  qualitative information on the 
 location and size of  localized states in a quantum Hall system
\cite{prl}, and will be reconsidered   in a forthcoming work.

 \section{  Magnetolocalization in disordered quantum wires}
 
 It is known  that the localization length 
 depends on the global symmetry of the wire~\cite{larkin}:
 $L_c = \beta \pi \hbar \nu S D_0  $, where $\beta =1, 2, 4$,
 corresponding to no magnetic field, finite magnetic field, and 
  strong spin- orbit scattering  or  magnetic impurities,
 respectively. $ \nu (E) $ is the electronic 
 density of states in  the wire\cite{reviews,ef}. 
 $ D_0 $ is the classical diffusion constant 
 of the electrons in the wire, and $S$ its crossection. 
  This result was obtained by calculating the spatial
 decay of the  density correlation function 
  for wires whose thickness exceeds the 
 mean free path $l$.

  Here, we 
 use an extension of a recent  nonperturbative calculation,  
 to obtain  the localization length 
 as a function of  the magnetic field, 
 using the fact that the ASD shows a crossover
from an oscillating behaviour, decaying 
with a power law\cite{haake,us}, typical for 
 Wigner- Dyson energy level statistics\cite{dyson}
 to a gaussian decaying function, when the length of the 
 wire is increased beyond the localization length\cite{prb}, 
 as seen in other measures of  correlations in the 
 discrete energy level spectrum of a phase coherent disordered 
 electron system\cite{ef,mirlin,guhr,altfuchs,isa} . 

 Taking the representation of the ASD derived above, Eq.
 (\ref{functionalintegral}), 
\begin{equation}
\bar{C}(\omega) = \int \prod d Q
 ({\bf x}) \exp( - F [ Q ] ), 
\end{equation} 
where 
the action Eq. (\ref{freeh} ) can   be rewritten conveniently 
in terms of the diffusion length, an electron
 would  diffuse classically in the magnetic phase shifting time $\tau_B$, 
 $L_B = \sqrt{ D \tau_B }$:
\begin{eqnarray}\label{freelb}
F[Q] &=& \alpha  \frac{1 }{16} L_{C  U}
 \int_0^L \mbox{d}  x \mbox{Tr} \left[ ( { { \nabla_x}} Q({ x}))^2
- \frac{1}{4 L_B^2} [ Q, \tau_3]^2 \right] 
 \nonumber \\
 &+& i \alpha \frac{\pi}{4}  \frac{\omega}{\Delta}
\int \frac{\mbox{d}  x}{ L} \mbox{Tr} \Lambda_3  Q({ x}).
\end{eqnarray}
where $L_{C U} = L_C (\beta = 2) = 2 \pi \hbar \nu S D_0$ is the
localization length in the wire in a moderately strong magnetic field 
\cite{larkin}.

 In the limit when 
 $ L_B < L_C $,  a moderately strong 
  magnetic field, $Q$  is reduced to a 
 $2 \times 2$- matrix by the   broken 
 time reversal symmetry.
  This reduces the space of Q to $ U(2)/(U(1) \times U(1) )$.

  For $\omega/\Delta  <  L_{C U}/L$, 
 corresponding to $ \omega < E_{\rm C}$, where 
$ E_{\rm C} = 2 \pi D/L^2$ is the Thouless energy scale of 
 classically free diffusion through the wire of length $L$,   
 the spatial variation of $Q$ 
 can be neglected 
  and one retains the same  ASD 
 as  for random matrices 
 of orthogonal or unitary symmetry, respectively \cite{haake,us}.

 Increasing the length of the wire $L$, a crossover 
 in the autocorrelation function can be seen as the wire 
 exceeds the length scale $L_c$\cite{prb}. 

  In order to study quantum localization along the wire,
 the function $C(\omega)$ should be thus considered 
as a function of the finite length  L of the  wire
 and spatial 
 variations
 of $Q$ along the wire have to be considered, as described by 
 the one dimensional nonlinear sigma model derived above. 
 
The 
 impurity averaged 
 ASD can to this end  be written as a partition function~\cite{prl}
\begin{equation} \label{part}
\bar{C} ( \omega ) = \mbox{Tr} \exp (
  - L  \bar{H} \left[ Q  \right]),
\end{equation}
  where $\bar{ H}$ is an  
 effective Hamiltonian of  matrices  Q
 on a compact manifold, determined by  
  the symmetries of the  Hamiltonian $H$ of disordered electrons.
 Thus, the problem  reduces to the one of finding the 
  spectrum of the effective Hamiltonian $\bar{H}$.

 We can  derive the corresponding Hamiltonian $\bar{H}$
 by means of the transfer matrix method, 
 reducing the one-dimensional integral over matrix field $Q$,  
Eq. (\ref{functionalintegral}),   to a single functional integral. 
 Thus, the ASD is obtained in the 
 simple form of Eq. (\ref{part}),
  with  the effective Hamiltonian 
\begin{equation} \label{effha}
\bar{H}( \omega = 0) =  \frac{1}{\alpha L_{CU} } (- 4 \Delta^R_Q
- \frac{1}{16}  X^2 Tr_Q [Q,\tau_3]^2).
\end{equation}
  $\Delta^R_Q$ is that
 part of the Laplacian on the  
 symmetric space, which does not commute with  $Tr[\Lambda_3 Q ]$.
 The time reversal symmetry breaking  due to the external 
 magnetic field is governed by the parameter  
 $X = \alpha  L_{CU}/( 2 L_B)$.

The problem is now  equivalent to a particle with ``mass''
   $ (\alpha/8) L_{CU} (E) $ moving on the symmetric space 
of $Q$ in a harmonic potential 
 with ``frequency'' $  1/( 2 L_{B})$,
and in an 
 external field $i \alpha (\pi/4) \omega/(L \Delta)$,  in ``time'' $x$,
 the coordinate along the wire.
To find the ASD as a function of $\omega$ and the length of the wire 
 $L$, one can do a Fourier analysis in terms of the spectrum and eigenfunctions
 of 
 the effective Hamiltonian at zero frequency, $\bar{H} ( \omega =0 )$
\cite{zirnbauer2}.

 There is 
   a finite gap $E_G$ between  the ground state energy and the 
 energy of the next excited state of  $\bar{H} (\omega = 0) $.
  For a long wire, $ L E_G \gg 1$, the ASD becomes,   
$C(\omega) = \exp ( - const. L \omega^2/E_G)$, 
 where both 
$ const. \omega^2 = \mid <~0~\mid \bar{H} ( \omega ) - 
 \bar{H} (0) \mid~1~> \mid^2$, 
 and the gap between the ground state and the first excited state, 
 $E_G = E_1 -E_0$  do   depend on the symmetry 
 of the Hamiltonian $\bar{H}$.
 This  exponential decay with 
 $ L\omega^2$  is  typical for a  a spectrum of localized states\cite{prl}.
 In the other limit $L E_G \ll 1$, 
all modes of $\bar{H}$ do contribute to the trace in the partition function 
Eq. (\ref{part}) 
with equal weight, yielding  the correlation 
 function of a spectrum of extended states\cite{prb}.
Thus, the crossover  length  is entirely determined by 
 the gap $E_G$, through 
$\xi_c = 1/E_G$, 
 and   can be identified with an averaged localization length.

 In order to derive the eigenvalues of the effective Hamiltonian at zero
 frequency, $\bar{H} ( \omega =0 )$, we need to introduce a representation 
 of the matrix $Q$ and evaluate
 the Laplacian in its parameters. This is done in Appendix C. 

  Without magnetic field, $B= 0$, 
 the Laplacian  is obtained to be 
\begin{eqnarray} \label{orth}
 \Delta^R_Q & = & \partial_{\lambda_C} 
 ( 1 - \lambda_C^2 ) \partial_{\lambda_C}
 + 2 \frac{1 - \lambda_C^2}{\lambda_C}   \partial_{\lambda_C}
\nonumber \\ 
& + & \frac{1}{\lambda_C^2}  \partial_{\lambda_D} 
 ( 1 - \lambda_D^2 ) \partial_{\lambda_D},
\end{eqnarray}
where $\lambda_{C,D} \in [-1,1]$.
 Its ground state is $1$ and the first excited state is $ \lambda_C
 \lambda_D$.
 Thus, the gap is 
\begin{equation}
E_G ( B = 0 ) = 16/L_{©C U}.
\end{equation}
 For moderate magnetic field, with the condition  
$ L_{C U} ( < y \bullet  y>  )^{1/2} B  \gg  \phi_0 = h/q$, all
  degrees of
 freedom  arising from time reversal invariance
 are frozen out,  due to the term  $Tr_Q [ Q, \tau_3]^2 = 16 (\lambda_C^2-1)$
 which fixes 
 $\lambda_C^2 = 1$. Then, the Laplacian reduces
 to 
\begin{eqnarray} \label{unit}
 \Delta^R_Q =  
\partial_{\lambda_D} 
 ( 1 - \lambda_D^2 ) \partial_{\lambda_D}.
\end{eqnarray}
 Its eigenfunctions are the Legendre polynomials. 
 There 
 is a  gap above the isotropic ground state
  of  magnitude
\begin{equation}
 E_G ( X \gg 1 ) =8/L_{C U}.
\end{equation}
   For moderate magnetic impurity scattering,
 exceeding the local level spacing,
 $1/\tau_s > \Delta_C$, $\alpha=2$,  and  the Laplacian is 
 given by Eq.(\ref{unit}).
 
   Thus, due to $\alpha=2$, the gap is   reduced
 to $
E_G (1/\tau_S > \Delta_C) = 4/L_{C U}$.
 For moderately strong 
spin- orbit scattering $1 / \tau_{SO} > \Delta_C$,
 the Laplace operator is 
\begin{equation} \label{spo} 
\Delta^R_Q =\sum_{l=1,2} \partial_{\lambda_l} 
 ( 1 - \lambda_l^2 ) \partial_{\lambda_l},
\end{equation}
 where $\lambda_{1,2} \in [-1,1]$. 
 The ground state is $\psi_0 = 1$,  the first excited state
 is doubly degenerate, $\psi_{11} = \lambda_1$, 
 $\psi_{12} = \lambda_2$.
  Thus, the gap is the same as for magnetic impurities,
\begin{equation} 
 E_G ( 1/\tau_{SO} > \Delta_C ) = 4/L_{C U}.
\end{equation}
 An external  magnetic field lifts 
 this degeneracy but does not  change the gap.
 
\begin{table}[bp]
\label{table}
\caption{
 Relation between symmetry
of the Hamiltonian 
 and the  gap of the quasi-1D- NLSM }
\begin{tabular}{|c|l|c|c|c|c|}
 Class & \multicolumn{2}{|c|}{ Symmetry  }
    &  Symmetric Space & Cartan class   & Gap  $E_G$  \\
\hline
Ordinary & T R  & S R        &   Sp(2)$/$( Sp(1) $\times$ Sp(1))
 & CII & $16/L_{C U}$       
 \\ \hline 
 Ordinary & No T R    &  S R  &  $ U(2)/(U(1) \times U(1) )$ ( Sphere )       & AIII &  $8/L_{C U}$         \\
\hline
Ordinary & T R  &  No S R     &    $ O(4)/(O(2) \times O(2) )$
 & BDI  & $4/L_{C U}$   
 \\ \hline
Ordinary & No T R &     No S R      &  $ U(2)/(U(1) \times U(1) )$      
& AIII &    $4/L_{C U}$     
\vspace{-.02cm}
\end{tabular} 
\end{table}

 Thus,
 using  the crossover in energy level statistics
 as the definition of a localization length 
 as  above, we get 
in a quasi- 1 -dim. wire,
\begin{equation} 
\xi_c = 1/E_G(\beta)= ( 1/16 ) \beta L_{C U} ,
\end{equation}
 where $\beta =1, 2, 4$
 corresponding to no magnetic field, finite magnetic field, and 
  strong spin- orbit scattering  or  magnetic impurities,
 respectively.  Comparing with the known equation for the 
localization length, $ L_c $, we find that 
 the dependence of the ratios  $\beta$ on the symmetry  are 
 in perfect 
agreement with the result as obtained from the 
 spatial decay of the density- density- correlation function\cite{larkin},
 while it defers by the overall constant $ 1/ 8$.
\begin{figure} \label{fig2}
\includegraphics[width=0.44 \textwidth]{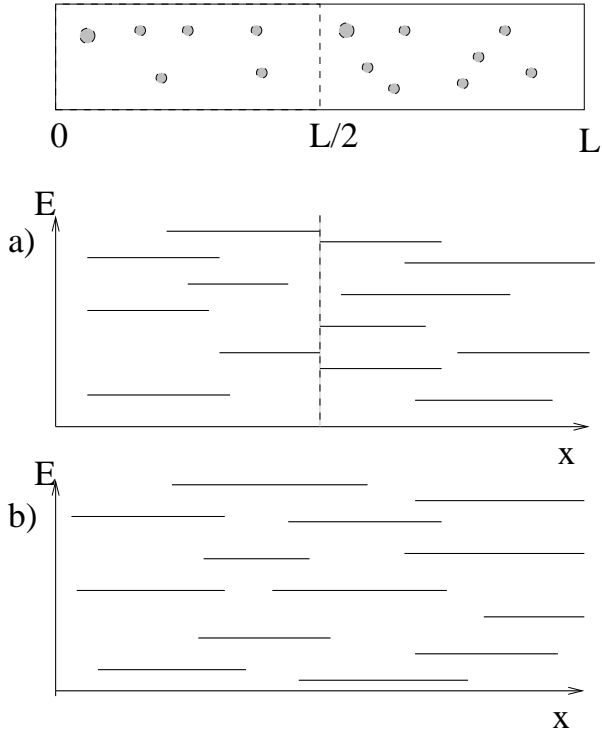} 
\begin{center}
\caption{ Schematic visualization of the
 energy level spectrum of localized states in a) 
 a disordered quantum wire of length L, 
  when divided into two parts, 
b) for 
  the same wire 
 when both parts are connected and the eigenstates  are
 hybridized. 
    }
\end{center}
\end{figure}

This relation can be proven directly. 
   The ASD at zero frequency  $\bar{C} ( 0 )_{L}$ of the wire of 
 length $L$, 
 becomes, when the wire is divided  into two parts,
 $\bar{C} ( 0 )_{L/2}^2$.
 For $L \rightarrow \infty$, we find that the relative difference
 is:
\begin{equation}\label{specdec}
f(L) = \frac{\bar{C} ( 0 )_{L/2}^2}{
\bar{C} ( 0 )_{L}}- 1  = 2 p \exp ( - L E_G/2 ), 
\end{equation}
exponentially decaying with the length $L$.
 Here $p$ is the degeneracy of the first excited state of
 $\bar{H} ( \omega = 0)$.  
  $f(L)$ can be estimated, following an argument by Mott\cite{mottdavis}: 
 When the two halves of the wire get connected,
 see Fig. (2), 
 the Eigenstates of the 
 two separate halves become hybridized and the Eigenenergy 
 of a state $ \psi_n $ is changed
 by $ \pm \Delta_C \exp ( - 2 x_n/L_C ) $. 
 $x_n$ is random, 
 depending on the position of an eigenstate with closest energy in 
 the other half of the wire. 
  Thus, averaging over $x_n$  gives:
\begin{equation} 
f( L  ) \sim + \exp ( - 4 L/ L_C ).
\end{equation}   
 Comparison with Eq. (\ref{specdec}) yields indeed $ 1/L_C = 8 E_G$.
 
  It is thus a remarkable fact that this length scale,  
 defined as the crossover length of the spectral autocorrelation 
 function, and related to the  excitation gap 
 of the compact nonlinear sigma model,  has exactly the 
 same symmetry dependence as the localization length,  defined through the 
 exponential decay of the spatial  density correlation function, 
 found in Ref. \onlinecite{larkin}. 
 This is   especially
 surprising,  since the nonperturbative 
 derivation   of the disorder average of the quantity, 
  $ < \rho ( {\bf r}, t ) \rho ( {\bf r'}, t' ) > 
 -  <\rho ( {\bf r}  )^2 >$,  necessitates the 
 use of the supersymmetry method, resulting in a nonlinear sigma model 
 of supermatrices, having 
 in addition to a compact sector, 
 the one considered here, a non compact sector, where the 
 matrix is parametrized on a  semi infinite interval. The full supersymmetry
 allows furthermore rotations between this compact and noncompact sector 
 which are parametrized by Grassmann numbers $\xi$, having the property
 $\xi^2= 0$. 
   Apart from this increase of the manifold of the matrix fields $Q$
 to the supersymmetric space, the structure of the theory is 
 equivalent. Especially, the free energy of the supersymmetric nonlinear
 sigma model, has exactly the same form as Eq. (\ref{freelb}), 
 replacing $ Q $ by supermatrices and the Trace over Q, by a supertrace
 $STr$, giving the opposite sign to the noncompact sector\cite{ef}. 

 Studying localization in a wire with this supersymmetric nonlinear
  sigma model,
 the transfer matrix method
 yields an effective Hamiltonian of supermatrices $Q$,
 of the same form as  Eq. (\ref{effha}), 
 where the Laplacian is now defined on the
 respective  supersymmetric  manifold. 
  In full analogy,  the spectrum of $\bar{H}$ determines
 accordingly  the properties of a disordered quantum wire, 
 and has  been derived in Ref. \onlinecite{zirnbauer2} for the pure 
 ensembles. 
  The partition
 function 
 $ Z = STr \exp ( - L \bar{H} ) $ is a generating function of 
 spectral correlation functions\cite{andreev,guhr}. 
  In order to derive spatial correlation functions like the 
 density correlation function, in addition, the 
 Eigen functions of the 
 respective diffusion  equation on the supersymmetric manifold, 
\begin{equation}
( - \partial_x + \bar{H} (Q) ) \psi(x; Q) = 0, 
\end{equation} 
  have to be found\cite{larkin}. 
  In that way, a formula for the conductance of a finite 
 disordered wire attached to two leads at a distance $L$, has been 
 derived\cite{zirnbauer2}, see also Ref \onlinecite{ef}. 
 In the limit of  a wire which is  perfectly coupled to the leads, 
 that formula for the average conductance simplifies to 
 \begin{equation}
 < g > = \frac{1}{2 \alpha} \int d \mu ( {l_i}  ) E ( {l_i}  ) 
\exp  ( - \frac{L}{16 }  E ( {l_i}  ). 
\end{equation}
  Where $  E ( {l_i}  ) $ are the eigenvalues of the supersymmetric
  Hamitlonian  
$\bar{H}( \omega = 0 ) $ and 
   $ d \mu ( {l_i}  )$ the corresponding integration measure, 
 of the discrete and continous eigenvalues of the 
 angular momentum operator on the compact and noncompact sector,
  respectively. 
 They were found to be given  for $B=0$ by\cite{zirnbauer2} 
\begin{equation} \label{spectrum1}
  E ( {l_i}  ) = 0, 
 4/L_{C U} 2 ( \epsilon^2 +1 ), 
  4/L_{C U} ( l^2 + \epsilon_1^2 + \epsilon_2^2 + 1),
\end{equation} 
 where $ l = 3,5,...$, and $ \epsilon > 0, \epsilon_1 > 0, \epsilon_2 > 0$. 

 For time reversal symmetry broken wires 
 $ X > 1$ the eigenvalues were found to be, 
\begin{equation} \label{spectrum2}
  E ( {l_i}  ) = 0,  
  \frac{4}{\alpha L_{C U})} ( l^2 + \epsilon^2 ),
\end{equation} 
where $ l = 1,3,5,...$, and $ \epsilon > 0$.

 If spin symmetry is broken, but  time reversal symmetry conserved, 
  in the presence of  spin orbit scattering,  
 the eigenvalues were found to be,
\begin{equation} \label{spectrum3}
  E ( {l_i}  ) = 0, 
 \frac{4}{ 2 L_{C U}} ( 2 (l-1)^2, 
 \frac{4}{ 2 L_{C U}}   ( l_1^2 +  l_2^2 +\epsilon^2 - 1),
\end{equation} 
 where $ l = 3,5,...$, 
 $ l_i = 1,3,5,.., i=1,2$
and $ \epsilon > 0$. 

    In that case it can be seen that for  a distance between 
 the leads much exceeding  the localizaion length, $ L \gg L_{C U}$, 
 the conductance decays  exponentiallly, and that this is  
 entirely determined by 
   the compact gap $\tilde{E}_{\rm G}$, between the lowest angular momentum 
 eigenstates of the compact sector. 
 The integration over the continous eigenvalues of the noncompact sector, 
 leads only to a prefactor, decaying as a power of the length,  $ \sim
  1/L^{3/2}$. 
 Indeed, the gap between the ground state value $ E=0$ and the first excited
 state is seen from Eqs.
 (\ref{spectrum1},\ref{spectrum2},\ref{spectrum3}), to be 
 $\tilde{E}_G =  8/L_{C U} $ for $B=0$, 
  $\tilde{E}_G =  4/L_{C U} $ for $X > 1$,  $\tilde{E}_G =  2/L_{C U} $ for 
 magnetic impurity scattering, $\alpha=2$, and   $\tilde{E}_G =  2/L_{C U} $ 
  for moderate spin- orbit scattering, coinciding with the 
 symmetry dependence of the compact 
 gap  derived above.
  However, that coincidence might appear as  mere chance, since in 
 fact, the Laplacian of the supersymmetric matrix 
  $Q$ can not be written as a sum of the one of the 
 respective compact nonlinear sigma
 model, Eqs. ( \ref{orth},\ref{unit},\ref{spo} ), because the 
 metric tensor $ \hat{g} $ on the supersymmetric space
 contains mixed factors of compact and noncompact parameters. 
  Therefore, the discrete eigenvalues of $-\Delta_Q$, 
 are not  the Eigenvalues
 of the square of the angular momentum on a compact sphere\cite{zirnbauer2}.
 Only, in the limit of infinite   noncompact parameters
  does one recover the 
 respective Laplacian on the compact symmetric space, 
 Eqs. ( \ref{orth},\ref{unit},\ref{spo} ).

  Thus, having shown that the ASD yields the correct symmetry dependence
 of the localization length,  we can now use this approach  
 to get an  analytical
 solution for   the 
 crossover behaviour of the localization 
 length and  the local level spacing  as a magnetic field is turned on, 
 and there is no   spin- orbit scattering.
  While a self consistent  approach \cite{bouchaud}, 
 a semiclassical analysis\cite{lerner} and
  numerical studies\cite{leadbeater,crossover}  showed  
 a continous increase of the localization length, an  analytical result
\cite{kolesnikov} indicated 
 that both limiting localization lengths
 $L_c(\beta = 1)$ and $L_c(\beta =2)$ are present in the crossover regime
 and that there is no single parameter scaling.
 This is explained by  arguing that the far tails 
 of the wavefunctions do cover a large enough area to have fully broken 
 time reversal symmetry, decaying with the length scale $L_c (\beta =2)$
 even if the magnetic field 
 is too weak to affect the properties of 
 the  bulk of the 
 wavefunction, which  does decay at smaller length scales with the shorter
  localization length $L_c (\beta =1)$, corresponding to the 
 time reversal symmetric case.
The quantity studied there is the imprurity averaged correlation function 
 of local wavefunction amplitudes and its momenta 
 at a fixed energy $\epsilon$:
$Y(\epsilon) = < \sum_{\alpha} \mid \psi_{\alpha} (0) \mid^2 
\mid \psi_{\alpha} ( r) \mid^2 \delta (\epsilon - \epsilon_{\alpha} )>$.
 It is   averaged  over  a distribution 
 of eigenfunctions in different impurity representations. Thus,  
  each eigenfunction could decay exponentially 
 with a single localization length, but having  a  distribution 
 which has two maxima, at  $L_c(\beta=1)$ and $L_c(\beta=2)$, 
 whose weight is a function of the magnetic field in the crossover regime.
 While the distribution function of $ \ln (\mid \psi_{\alpha} (0) \mid^2 
\mid \psi_{\alpha} ( r) \mid^2 ) $
 is known to be Gaussian in both limiting cases of 
conserved, and fully broken time reversal symmetry, centered 
 around the value $  r/L_{C } ( \beta ) $, $ \beta = 1,2$, respectively,  
 it is not yet known
 in this crossover regime, 
 however\cite{mirlin}. The average value of 
 moments,   $\mid \psi_{\alpha} (0) \mid^k 
\mid \psi_{\alpha} ( r) \mid^k$, is decaying  more slowly than its 
 typical value, and does not depend on the order of the moment,$k$.  
 This was taken as a proof that moments are determined by states with
 anomalously large localization lengths of the order of the system
 size\cite{mirlin}. 
 Therefore, 
the result of Ref.  \onlinecite{kolesnikov}
 can  be  a property of such rare states with  anomalously large 
 localization length, and it remains to see, if the full distribution 
 function scales with two lengths $ L_c ( \beta )$, $\beta = 1,2$, 
 or a single one, changing continously with 
 the magnetic field, $ L_c ( B ) $. 

 While we cannot  resolve this question by calculating a
 spectral autocorrelation function like the  ASD, 
 this is another motivation  to see if the energy level statistics 
  is  governed by a single parameter as the magnetic 
field  is varied. 
 
 The effective Hamiltonian 
 for  moderate magnetic fields is found, without spin dependent scattering,
 $\alpha=1$, using $ Tr [Q, \tau_3 ]^2 = 16  ( 1-\lambda_C^2)
$  to be given by:
\begin{equation}
\bar{H} =\frac{1}{L_{C U}} ( - 4 \Delta^R_Q + X^2 ( 1-\lambda_C^2) )  ,
\end{equation}
 where the Laplacian is 
 Eq.(\ref{orth}) and 
  $X =  L_{C U}/(2 L_B) $.
 
In the limit $ X \rightarrow 0$ the ground state and first 
 excited state  approach $ 1, \lambda_C \lambda_D$, respectively.
In the limit $ X \gg 1$, $\lambda_C^2$ becomes fixed to 1.
 Thus, the 
Ansatz $\psi_0(\lambda_C ) \sim \exp ( A_0 X^2 (1- \lambda_C^2))$,
 and $\psi_1(\lambda_C,\lambda_D ) \sim \lambda_C \lambda_D \exp ( A_1
 X^2 (1-\lambda_C^2))$,
 where $ A_0 < 0, A_1 < 0$ are negative constants,
 solves $\bar{H} \psi =\bar{ E} \psi$
 to first order in  $z= X^2 ( 1-\lambda_C^2)$. 
 One finds that the two lowest  magnetic field dependent eigenvalues
 are
$E_0 = 4/L_{C U}  ( -5 + \sqrt{ 25 + X^2} )$, and 
 $E_1 = 4/L_{C U}  ( -3 + \sqrt{ 49 + X^2} )$, 
 and  the Eigenfunctions are given as above with 
 $ A_0 = - L_{C U} E_0/ (16 X^2) $, and $ A_1 = ( 1 - L_{C U} E_1/ 16 )/X^2$, 
 yielding the right limits for
  $ X \rightarrow 0$ and $ X \gg 1$, respectively. 
 Thus,  there is a magnetic field dependent gap
$ E_G = E_1 -E_0 $
 of magnitude:
\begin{equation} \label{smooth}
E_G (X) = 4( 2 + \sqrt{49 +  X^2} - \sqrt{25 +  X^2})/L_{C U}.
\end{equation}
 This  solution is valid in both 
 the limits $X \ll 1$ and $X \gg 1$, interpolating 
  the region $ X \approx 1$.

 With the magnetic diffusion length $L_B = ( D \tau_B )^{1/2}$, 
 and the magnetic phase shifting rate,  as given by
 Eq. (\ref{magneticphaseshifting}), we obtain: 
\begin{equation}
 X = L_{C U}/ (2 L_B) =  L_{C U} \frac{q}{ \hbar}   \sqrt{ < y \bullet  y> } B,
\end{equation}
 which is 
$  \sqrt{ < y \bullet  y> }/ W$ times 
 the number of flux quanta penetrating
 a localization area
 $ L_{C U} W  $.

 From Eq. (\ref{smooth}) follows that the 
  magnetic  change of  the
 localization length  is $\delta L_C(B) \sim B^2$ for small  
 and $\sim 1/B$ at large magnetic fields,  
 which  agrees with the result of the selfconsistent method
 as obtained 
 by Bouchaud\cite{bouchaud}.

\section{Resistance of disordered  Quantum Wires} 

 In the limit of zero temperature, 
 $T=0$,  the resistivity of a disordered quantum wire, having only  localized 
 states at the Fermi energy,  is infinite. 
 For finite temperature,   $T>0$, 
  in the strong localization regime 
 $k_{\rm B} T < \Delta_{\rm C}$,  the 
mechanism of conduction is hopping of electrons between localized states.
 Then,  the resistivity increases exponentially with temperature. 
According to the resistor network model\cite{Miller,Shkl1}, each pair of 
localized states $i$ and $j$ is linked by a resistance $R_{ij}$:
\begin{equation} \label{resistor}
R_{ij}=\exp(\frac{2r_{ij}}{L_{c}}+\frac{\epsilon_{ij}}{k_{B}T})
\end{equation}
where $r_{ij}=|r_{i}-r_{j}|$ and \(\epsilon_{ij}=(|\epsilon_{i}-\mu|+
|\epsilon_{j}-\mu|+|\epsilon_{i}-\epsilon_{j}|)/2k_{B}T\) ($r_{i}$ and 
$\epsilon_{i}$ are the position and energy of the state $i$, $\mu$ being the 
Fermi energy).
Because of the exponential dependence of $R$ on $r_{ij}$ and 
$\epsilon_{ij}$, percolation theory methods can be  applied
\cite{Ambeg,Pollak,Shkl2}. In 2-D and 3-D systems, 
 the dependence 
of $R$ on  temperature 
  $T$ shows a crossover from an activated behaviour to the 
variable range hopping (VRH) regime. In this regime the temperature is so low 
that the typical resistances between neighbouring states are large because of 
the 
 second term in Eq. (\ref{resistor}).
 Therefore electrons tunnel to distant states whose energies 
are close to the Fermi level.
If we neglect electron-electron interactions the resistivity is described by 
Mott's law \cite{Mott,Ambeg}:
\begin{equation} \label{mott}
R(T)=R_{0}\exp[(\gamma T_{0}/T)^{1/(d+1)}] 
\end{equation}
where $d$ is the dimensionality of the system, $\gamma$ a numerical coefficient
which depends on $d$, $T_{0}=1/ \nu L_{c}^{d}$ and $\nu_d$ is the
 dimens dependent   density of 
states.
However, in the quasi-1-D 
case and for sufficiently long wires 
 the variable range hopping result,  Eq. (\ref{mott}), 
 cannot used  due to the presence of exponentially rare 
segments inside which all the localized states have energies far from the 
Fermi level \cite{Lee,Raikh1,Kurki}.
These large resistance segments (LRS) do not strongly  affect 
  the resistivity of 2-D 
and 3-D systems because they can be circumvented by the current lines. In 1-D 
this is not possible and the total 
resistance of a wire is given by the sum of the resistances of all the LRS's.
This sum yields an  activated type dependence of $R$ on $T$\cite{Raikh1}
 for infinite wires: 
\begin{equation}\label{raikh}
R=R_{0}\frac{L}{L_{c}}(\frac{T_{0}}{T})^{\frac{1}{2}}\exp(T_{0}/2T),
\end{equation}
where $k_{\rm B} T_{0}=1/ \nu L_{c} = \Delta_{\rm c} $
 coincides with the local level spacing, 
 and $L$ is the length of the wire.
 Eq. (\ref{raikh}) 
 is valid provided that the number of optimal LRS's (i.e. those 
LRS's which give the largest contribution to $R$ \cite{Raikh1}) within
 the 
 length of the sample) is large.
 Bur for a finite wire length this condition fails to be fulfilled
at very low temperature $T$,  
 and the resistance of the
chain is determined by smaller LRS's; in this regime Eq. (\ref{raikh})
 is 
replaced by  \cite{Lee,Raikh1}:
\begin{equation}
R\approx R_{0} \exp  [ \sqrt{ 2\frac{T_{0}}{T} 
\log(\frac{L}{L_{c}}(\frac{T}{T_{0}})^{\frac{1}{2}}\log^{\frac{1}{2}}(\frac{L}{L_{c}}))
}
],
\end{equation}
which is valid below a temperature 
\begin{equation}
T_{1} =\frac{T_{0}}{2 \ln(L/L_{c})},
\end{equation}
 approaching Mott's law, Eq. (\ref{mott})  at  lower temperatures 
 $T < T_1$.

 So far, 
 electron-electron interactions have not been taken into 
account.  This approximation is valid if the Coulomb interaction is  
screened over distances of the order of the hopping length, 
 as by  a metal 
gate electrode  deposited on top of the wires at a distance smaller than 
the typical hopping lengths. When this is not the case, 
 long range  electron-electron interactions 
affect both the density of states and the resistance of the samples
\cite{Raikh2,Larkin}.

\section{Comparison with experimental results}

 The magnetic field dependent 
 activation energy was
 measured recently in  transport experiments
 of Si $\delta$- doped Ga As quantum wires\cite{khavin}.
 As an example, we discuss here the sample 5 of Ref. \onlinecite{khavin}, 
 with a width $ W = .2 \mu m$, a  
 localization length $L_{CO} = .61 \mu m$ a length $ L = 40 \mu m$, 
 and $ N = 30$ channels.

  The activation energy coincides with the 
 local  level spacing  $k_{\rm B} T_0 =  \Delta_c = 1/ (\nu W L_c) $ 
 and is estimated for sample 5 to be $ T_0 = .34 K$. 
 
 Thus, according to the theory outlined in the previous 
 section, there is an activated
 reistance in an order of magnitude  temperature range 
 $ T_1 =   .04 K < T < T_0 = .34 K$, allowing 
 in good approximation the direct measurement of the 
 magnetic field dependent 
 activation energy $  \Delta_c ( B)  $, and thus the
 magnetic field dependence of the 
 localization length 
 $L_C ( B) $.  
 
 The  ratio 
 of the cyclotron frequency and the elastic scattering rate,
 $ \omega_C \tau = l/(k_{\rm F} l_B^2) \ll 1$, is small  in the whole 
range of magnetic fields considered there, so that the classical 
 conductance would be magnetic field independent, 
 $\sigma = n e^2 \tau/m (1 +\omega_C^2 \tau^2)^{-1} \approx n e^2 \tau/m$. 

 The mean free path $l \sim .02 \mu$ is small compared to the 
 width of the sample $W = .2 \mu m$.  The magnetic length is 
 $l_B = .026 \mu m ( B/T )^{-1/2}$. 
 Thus,  while $ \omega_C \tau \ll 1$, the magnetic length 
 becomes smaller than the width of the sample at magnetic fields 
$ B > .0165 T$. 

The experimental magnetic field dependence 
 of the ratio of activation energies is shown 
 in Fig. (3) 
 together with the theoretical curve for the 
 ratio of local energy level spacings 
 $\Delta_C(B)/\Delta_C(0) = E_G(B)/E_G(0)$,
 as derived above, 
 Eq. ( 36 ), 
using for the 
 magnetic phase shifting rate the results for a 
 2-dimensional wire with specular boundary conditions, 
 Eq. (14), and, for comparison,  the one derived for a parabolic wire, 
 Eq.  (\ref{parabol}).  

 There is a quantitative discrepancy 
 between the best fit  $X = .036 B/G$,  
 and 
  $X =  2 \pi \phi/\phi_0 $, 
  $\phi = \mu_0 H L_{C U} (\overline{y^2})^{(1/2)}$, when using the 
 analytical formula  Eq. ( 14).
 With the experimental 
 parameters   $ \alpha=1, L_{CO} = .61 \mu m$, width $W = .2 \mu m $ 
of  sample 5 in Ref. \onlinecite{khavin} 
 and   $ \overline{y^2}  = W^2/12$ for a 2- dimensional wire, it yields
 rather  $ X= .010 B/G$. We note that smooth confinement can give  
 $\overline{y^2}  > W^2/12$. 
 A similar discrepancy was  observed  between $W$ as obtained 
 from the sample resistance
 and estimated from the analysis of the weak localization magnetoresistance,
 which also depends on $\overline{y^2}$\cite{khavin2}.

 We note that the agreement, when using the
 experimental parameters,  
 for   the parabolic wire,  is
 better.   
  The cyclotron length
  $ l_{\rm cyc} = k_{\rm F} l_{\rm B}^2 = .32/(B/T) \mu m$, is found to be 
  larger than the mean free path $l$ for $B < 15 T$ and 
 larger  than the
 wire width  for $ B < 1.5 T$. 
 We find for the parabolic wire: 
 $X = .024 ( .99 + 1.33~10^{-8} ( B/G)^2 )^{1/2} B/G $. 
  The enhancement of the magnetic phase shifting rate 
 in a parabolic wire, Eq. (\ref{parabol}), is thus too weak to be 
  seen at the magnetic fields 
 used in the experiment,  $B < .2T $, as shown in  Fig. (3), 
 and seems thus not to be the origin of the 
 increase in the  decay of the activation gap, at about $.1T$ .

  An extension 
 of the derivation given in section IV to include a dependence
 of the eigenfunctions on the magnetic field
 also for a  2-dimensional wire with specular boundary conditions
 has to be done, in order to make the comparison with  the experiment 
 more quantitative, and conclude from the
 magnetolocalization on the form of  the 
 confinement potential in these   Si- $\delta$- doped Ga As quantum wires.
  But, our results may indicate that the harmonic confinement model 
 of the parabolic wire is a better description of the wires in 
 sample 5. 

\begin{figure} \label{fig4}
\includegraphics[width=0.44 \textwidth]{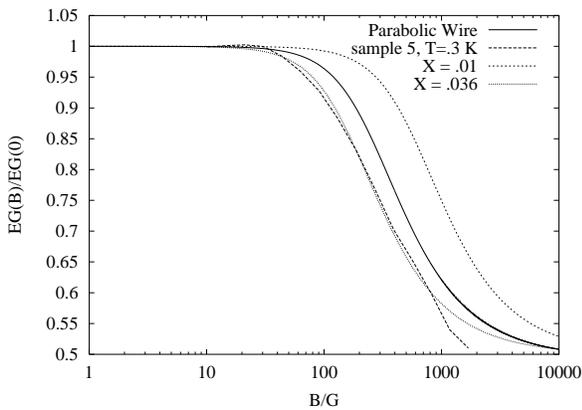}
\begin{center}
\caption{  The  activation gap ratio
  $T_0(H)/T_0(0)$ as a function of the magnetic field $B$ in $G$
of sample 5 measured at temperature $T= .3 K$ as reported in Ref. 1, 
 together with the theoretical curves for a parabolic wire, 
 using the parameters of sample 5.  and 
 a   2D wire with specular boundary conditions
 for a best fit value  $ X= .036 B/G $, and the value obtained
 from the experimental parameters, 
 $X = .010 B/G$.     }
\end{center}
\end{figure}

\section{ Summary and open problems}

\label{sec:summary}

   A formula for  the   magnetic phase shifting rate 
 has been derived, which allows its calulcation for arbitrary wire 
 geometries and ratios of the elastic mean free path,
 the wire width, and  
 the 
 magnetic length. 

 For a quantum  wire with  specular boundary conditions and  harmonic 
 confinement
 this formula has been evaluated explicitly, and compared with 
 previous analytical and numerical results for the
 magnetic phase shifting rate.  
 
 The 
  localization length is 
 derived as the crossover length scale 
 from correlated to uncorrelated energy 
 level statistics, as studied with the autocorrelation function 
 of spectral determinants.
   It is shown that its symmetry dependence
 coincides exactly   with the 
 localization length as defined by the 
 exponential decay of  the averaged two-terminal conductance
 and  derived with the supersymmetry method. 

 Therefore, the ASD can be  used to get analytical information on  
 the magnetic field dependence of the localization length, 
 which is shown to be governed by the  magnetic phase shifting rate, 
 and thus strongly dependent on the geometry of the wire and the 
ratios of the elastic mean free path,
 the wire width, and  
 the 
 magnetic length.

 A  comparison with the magnetic field dependence of the 
 activation gap, as  observed in 
  low temperature resistance measurements 
 in Si $\delta$- doped Ga As wires,  
 indicates,  that the electrons  move in a potential which is 
  closer to a 
  harmonic 
  than a   hard wall confinement.

 Enhancement of the sensitivity of the localization  to a magnetic
 field is found analytically,   when the 
cyclotron length is comparable  with 
 its  width. The physical reason for this 
 enhancement is found to be the magnetic field dependent shift  of the 
 guiding centers  of the electronic eigenstates in the quantum wire, 
 even at moderate magnetic fields, when the classical conductivity
  is still 
 independent of the magnetic field.

 It remains to extend the derivation to include random surface scattering 
 \cite{leadbeater}
 and the effect of correlated, smooth  disorder\cite{aleiner}, 
 in order to allow for  a more quantitative comparison with the experiment. 
  Both effects necessitate  a new  
 derivation of the nonlinear sigma model, which  
 allows for a directional dependence of the matrix field $Q$. This has been  
  recently introduced for a system with broken time reversal symmetry
 in the  study of localization in  correlated disorder 
\cite{taras1}, 
 and the  spectral statistics of quantum billards with surface
  scattering\cite{blanter}. 
 In both cases one is lead to a nonlinear sigma model, 
 where variations of the matrix $Q$ on ballistic length scales 
 are taken into account\cite{muzy,aasa,simons}.
  The application of this approach to the magnetolocalization in 
 disordered quantum wires will  be presented in a future publication.

\section*{ACKNOWLEDGMENTS}

The  authors gratefully acknowledge, 
 usefull discussions with Isa Zarekeshev, and Konstantin Efetov,  thank  
 Yuri Khavin for providing his data and 
usefull communications, and Bernhard Kramer for stimulating support
 and critical reading of the manuscript. 

\section*{Appendix A}

 Here, the derivation for spinless case, 
 $\alpha=1$, is given in detail. 
 We   use for
  compactness the vectors of anticommuting variables, 
\begin{equation}\label{psi}
 \psi ({\bf x}) = \left( \begin{array}{c} \xi ( {\bf x} )
 \\ \xi^* ({\bf x}) \\
\eta ( {\bf x}) \\ \eta^*( {\bf x} ) \end{array} \right),
 \bar{\psi} ( {\bf x} ) = ( \xi^* ( {\bf x} ), - \xi ( {\bf x} ),
 \eta^*({\bf x}), - \eta ({\bf x}) ). 
\end{equation}
 Note that $ \bar{\psi} = ( C \psi )^T $, 
 where the matrix  $C$ interchanges the Grassmann fields with their 
 conjugate one, and has thus the form 
$ C = \left( \begin{array}{cccc} 0 &  -1 & 0 & 0 \\  1 & 0 & 0 & 0 
\\ 0 & 0 & 0 & -1 \\ 0 & 0 & 1 & 0   
\end{array} \right)$. 

Thus, the ASD  is written as 
\begin{eqnarray} \label{grassmannrep}
\bar{C}(\omega ) &=&  \int \prod_{\bf x} d \psi({\bf x}) 
\nonumber \\ &&
\exp ( -  \frac{1}{2}\int d {\bf x} \bar{\psi} ({\bf x}) 
(E + \frac{1}{2}\omega \Lambda - \hat{H_0} - V({\bf x})  )\psi({\bf x}) ), 
\end{eqnarray}

  Here, the  diagonal Pauli 
matrix $ \Lambda = \left( \begin{array}{cc}  1 & 0 \\  0 & -1 
\end{array} \right)$ has been introduced for compactness, its diagonal 
 elements projecting on the respective spectral determinant of 
 the ASD. 
The kinetic  Hamiltonian becomes   a matrix 
\begin{equation}
 \hat{H_0} =
 (\hat{p}- q \tau_3 {\bf A}  )^2/2/m,  
\end{equation}
where 
 the  diagonal Pauli matrix 
$\tau_3$ had to be introduced  since each vector 
 has elements  of the Grassmann field and  the time reversed one, and
 the diamagnetic 
term ${\bf p A}/m$ 
in the Hamiltonian  changes sign, as  ${\bf p} \rightarrow 
-{\bf p}$, breaking the time reversal invariance.

 To summarize the notation, here, and in the following, 
$\Lambda_i$ are the  Pauli matrices in the subbasis
 of the left and the right spectral determinant,
$\tau_i$ the ones in the subbasis spanned by  time reversal 
and $\sigma_i$  the ones in the subspace spanned
 by the spinor, for  $i=1,2,3$.  

 Note that a global transformation  of the Grassmann vectors 
 $\psi \Rightarrow \tilde{\psi} = A \psi $ does leave the 
 functional integral for $\omega=0$
 invariant, as long as  $ A^+ A = 1$, 
and $A^{+ T} C = C A$, restricting the matrices $A$ to be symplectic
 ones, being  elements of $Sp(2)$,
 commuting with the antisymmetric matrix $C$. 
 A finite frequency breaks this symmetry group, 
 and only  symplectic transformations
 of each field of a single  spectral determinant 
 separately, $ Sp(1) \times Sp(1) $,  do leave the functional integral invariant. 
 
 Now, the 
 averaging over the disorder potential
 can be done, integrating 
 Eq. (\ref{grassmannrep}) over the   Gaussian distribution function 
 of the random potential V.   

 Thus, the averaged ASD is found to be given by 
 a functional integral over interacting Grassman fields 
 $\psi$, 
\begin{eqnarray}
\bar{C}(\omega ) &=&  \int \prod_{\bf x} d \psi({\bf x}) 
\nonumber \\ &&
\exp ( -  \frac{1}{2}\int d{\bf x} \bar{\psi} ({\bf x}) 
(E + \frac{1}{2}\omega \Lambda - \hat{p}^2/2/m )\psi({\bf x}) ) 
\nonumber \\
&& \exp(
- \frac{1}{16 \pi} 
 \frac{\hbar \Delta}{\tau} S L 
\int d {\bf x}  Tr (\psi({\bf x}) \times \bar{\psi}({\bf x}) )^2).
\end{eqnarray}
  
Now, the  resulting
$\psi^4$-interaction term can
 be decoupled by introducing 
another Gaussian  integral over 
an auxilliary field. Clearly, the field should not be a scalar, 
 otherwise we would simply reintroduce the Gaussian integral over 
 the random potential $V$.
 Rather, 
 in order to go a step forward,
  the auxilliary field 
 should capture the full symmetry of the autocorrelation function.
Therefore, the 
Gaussian integral is chosen to 
 be  over a 
4 by 4 matrix $Q_{4 \times 4}$,
 which is 
  itself an 
 element of the respective 
 symmetric space, as the matrix $A$ which  leaves 
  the functional integral invariant. 
 Thus, allowing for 
 a spatial dependence of $Q$,  one can decouple  the 
 interaction term: 
\begin{eqnarray} \label{hst}
&&\exp( -\frac{1}{16 \pi} \frac{\hbar \Delta}{ \tau} 
 S L \int d {\bf x} ( \psi({\bf x}) \times \bar{\psi}({\bf x}) )^2)
\nonumber \\
&=& \int \prod_x d Q_{4 \times 4}
({\bf x}) \exp( - \pi \frac{ \tau}{\hbar \Delta} \int
 \frac{d {\bf x}}{S L}
 Tr Q_{4 \times 4} ({\bf x})^2 
\nonumber \\
&+& 
i \frac{1}{2} \int d {\bf x} Tr Q_{4 \times 4} ({\bf x}) \psi({\bf x}) \times \bar{\psi}({\bf x})).   
\end{eqnarray}
 Anticipating, however, that the functional integral over 
 the matrices $Q$ cannot be performed exactly, but rather only 
an integral 
over slowly
 varying modes around a saddle point solution, 
 it is necessary to separate fast and slowly varying modes already 
 before the decoupling of the interaction term  
 Eq. (\ref{hst})\cite{elk}. 
  It turns out that there are two equivalent slowly varying 
 interaction terms, corresponding to diffusion, and 
  one arrives finally after a Gaussian   decoupling  to 
  a, by a factor $1/2$, shallower nonlinear coupling $ Tr Q^2$\cite{ef}.

 Next, one can perform the Gaussian
 integral over the Grassmann vectors $\psi({\bf x})$
 and  one   obtains  for the ASD, 
 rescaling $ Q_{4 \times 4} \rightarrow 2 \tau/\hbar Q_{4 \times 4}$, 
 the  representation:

\begin{equation}\label{functionalintegral}
\bar{C}(\omega) = \int \prod d Q_{4 \times 4}
 ({\bf x}) \exp( - F [ Q ] ), 
\end{equation} 
with 
\begin{eqnarray} \label{exactfree}
F[Q] = && \frac{\pi}{8}
 \frac{\hbar}{\Delta \tau}
  \int \frac{d {\bf x}}{S L} Tr Q_{4 \times 4}({\bf x})^2 )
\nonumber \\ &&
 + \frac{1}{2} \int  d {\bf x}
 <{\bf x} \mid Tr \ln  (
 G ( \hat{x}, \hat{p} ) \mid {\bf x} >,
\end{eqnarray} 
 where
\begin{equation}
 G( \hat{x}, \hat{p} ) = 1/( 
\frac{1}{2}  \omega \Lambda - 
\frac{(\hat{p}- q \tau_3 A)^2}{2 m} + i \frac{\hbar}{2 \tau} 
Q_{4 \times 4}(\hat{x}) ).
\end{equation}
 is the  propagator matrix. 
 We used the operator 
 notation $\hat{x}$, in order to stress that the terms in the inverse
 propagator, do not commute with each other.

\section*{Appendix B} 

 For  a clean wire with hard wall boundaries, the transversal 
 Eigen modes are 
 for $ -W/2 < y < W/2$, 
$< k_y \mid y > = \cos k_y y$ for
 $ k_y = \pi s/W$, $s$ being an odd integer, 
 and $< k_y \mid y > = \sin  k_y y$
 for 
 $ k_y = \pi  s /W$, s being an even  integer,  one obtains: 
\begin{equation}
\mid < k_y \mid y \mid k_y' > \mid^2 = \frac{1}{W^2} 
( \frac{1}{(k_y -k_y')^2} -  \frac{1}{(k_y  + k_y')^2})^2, 
\end{equation}
 when $ k_y = \pi s/W$, and  $ k_y' = \pi  s' /W$, 
 s being even, and s' odd, or vice versa. 
  Then, the sum over $k_y'$ 
 in Eq. ( \ref{h} ) can be performed by use of the
 Matsubara trick, for $s$ even,  and odd integers,  separately. 
 The remaining sum over $k_x,k_y$ can be transformed as
 $1/(W L) \sum_{k_x,k_y} = \int d \epsilon \nu ( \epsilon ) 
\int \frac{d \hat{e}_k}{\Omega_k} $, noting that the 
 unit vector $\hat{e}_k$ can point only in discrete directions. 
 Thus, while in 2 dimensions
 $ \int \frac{d \hat{e}_k}{\Omega_k} = \int_0^{2 \pi} 
 \frac{ d \theta }{ 2 \pi } = 4/(2 \pi) \int_0^1 d y 1/( 1-y^2)^(1/2) $,
  for finite number of transverse channels $N = k_F W/\pi$ 
 there  is a sum, 
$\int \frac{d \hat{e}_k}{\Omega_k} = 2/(\pi N ) \sum_{s > 0} 1/(1
 - s^2/N^2)^(1/2)$.
 Thus,  $k_y = \pi s/W= k_F s/N$ and $k_x = k_F ( 1 - s^2/N^2)^(1/2)$.
 Performing finally for $ E \gg \hbar/\tau$ the integral over $\epsilon$, 
 one arrives with some patience  at Eq. (\ref{hexact}), where 
 $K_0 = 2/(\pi N) \sum_{s=0}^{N}  \sqrt{1- \frac{s^2}{N^2} }$, 
$K = 2/(\pi N) \sum_{s=1}^{N}  \sqrt{1- \frac{s^2}{N^2} }$,
$K_1 = 2/(\pi N) \sum_{s=1}^{N}  \sqrt{1- \frac{s^2}{N^2} }/s^2$,
$K_2 = 2/(\pi N) \sum_{s=1}^{N}  \sqrt{1- \frac{s^2}{N^2} } s^2/N^2$.

\section*{Appendix C}

 In order  to derive  the Laplacian in the respective 
 representation of the matrix field $Q$,  its general definition 
 in an arbitrary parametrization,  
\begin{equation} \label{laplacebeltrami}
\Delta_Q = \frac{1}{\sqrt{ \bar{g} }} \sum_{i,k} \partial_k g^{ik} \sqrt{
\bar{g} }  \partial_i, 
\end{equation}
 where the matrix $g$ is  the metric tensor, being 
defined by the quadratic form 
 $ ds^2 = 1/4 Tr dQ^2$ of the representation 
\begin{equation}
d s^2 = d {\bf x}^T g  d {\bf x},
\end{equation}
 where $ {\bf x}$ is the vector of parameters of the
 representation.  

For 
{\it $B \neq 0 $}, $Q$  is element of $U(2)/ ( U(1) \times U(1) )$, 
 by enforcing the conditions $ Q^2 = 1$, $Q^T C = C Q$, and $Q^+ = Q$, 
 $ [ Q,\tau_3 ] = 0$. 

It can be paramterized as 
$ Q = \left( \begin{array}{cc}  \cos \theta  & e^{i \chi} \sin \theta
 \\  e^{-i \chi} \sin \theta  & - \cos \theta  
\end{array} \right)$

 where $ \theta \in [ 0, \pi ] $ and $\chi \in [0, 2 \pi] $. 

 Thus, 
\begin{equation}
d s^2 = d \theta^2 + \sin^2 \theta d \chi^2. 
\end{equation}
 and 
$ g = \left( \begin{array}{cc}  1  &  0
 \\  0  &  \sin^2 \theta  
\end{array} \right)$. 
 Thus, with Eq. (\ref{laplacebeltrami}) follows: 
\begin{equation}
\Delta_Q =  
\partial_{\lambda_D} 
 ( 1 - \lambda_D^2 ) \partial_{\lambda_D} + \frac{1}{1- \lambda_D^2} d \chi^2,
\end{equation}
 where $ \lambda_D = \cos ( \theta ) $. 
 
 Note that the autocorrelation function depends on the energy difference
 $\omega$ through the coupling 
 $Tr \Lambda Q = 2 * 2 \lambda_D$, 
  so that  only that part of the Laplacian which does not 
 commute with  $Tr \Lambda Q $, 
 \begin{equation}
\Delta^R_Q =  
\partial_{\lambda_D} 
 ( 1 - \lambda_D^2 ) \partial_{\lambda_D}.
\end{equation}
  enters in the
 frequency dependence of the 
  autocorrelation function of spectral determinants,  
 Eq. ( \ref{part} ).  
 Since $U(2)/(U(1) \times U(1) ) = S_2$, the two sphere, this is equivalent to 
 the treatment of spherically symmetric potentials, 
and the Laplacian can be identified with the square of the 
 angular momentum, $ - \Delta_Q = {\bf L}^2$, 
 and $ L_z = i \partial_{\chi}/( 1 - \lambda_D^2 )$
 does commute with the Hamiltonian, 
\begin{equation}
 \bar{H} = - 1/(2 m) L^2 + i \alpha \frac{\pi}{4} \frac{\omega}{\Delta} z,  
\end{equation}
  since $ z = \cos \theta_D $ does commute with $L_z$. 
 Therefore,  $ \omega \neq 0$   
does not break the azimuthal symmetry of rotations around the 
 z-axis, $n_z$.

 For {\it $ B= 0$},
 $Q$  is element of the symplectic symmetric 
 space, $Sp(2)/ ( Sp(1) \times Sp(1) )$, 
 by enforcing the conditions $ Q^2 = 1$, $Q^T C = C Q$, and $Q^+ = Q$. 

One obtains:
\begin{equation}
Q = \left( \begin{array}{cc} c \openone & A \\ A^+ & - c \openone
\end{array} \right).
\end{equation}
with $A= \left( \begin{array}{cc} a & b \\ b^* & -a^* \end{array} \right)$
where $ \mid a \mid^2 +
\mid b \mid^2 +  c^2 = 1$.

 A matrix Q with the above symmetries can be represented as,
\begin{equation}
Q= U^{-1} Q_c^0 U,
\end{equation}
with
\begin{equation}
U = V_C U_D,
\end{equation}
where
\begin{equation}
U_D = V_D^{-1} T_D^0 V_D,
\end{equation}
where
\begin{equation}
 Q_c^0 = \left( \begin{array}{cc} \cos \theta_C & i \sin \theta_C \tau_2
\\  i \sin \theta_C \tau_2  & - \cos \theta_C \end{array} \right),
\end{equation}
and
\begin{equation}
T_D^0 = \left(\begin{array}{cc} \cos \theta_D/2 & i
\sin \theta_D/2 \\ i \sin \theta_D/2 & \cos \theta_D/2
\end{array} \right).
\end{equation}
and
\begin{equation}
V_{C,D} = \left(\begin{array}{cc} \exp ( i \phi_{C,D} \tau_3 ) & 0
 \\ 0 & \openone
\end{array} \right).
\end{equation}
and $\tau_i, i=1,2,3$ are the Pauli matrices.  Such a representation
 was first given by Altland, Iida and Efetov\cite{altland} to study
 the crossover between the spectral statistics of Gaussian distributed random 
 matrices  as the time reversal symmetry is broken, 
 within the supersymmetric nonlinear
 sigma model. Here, in order to study the ASD, 
 we need to  consider only the compact block of the
 representation given there.

 We find that 
 $ ds^2 = Tr dQ^2/4 = d \theta_C^2 + \cos^2  \theta_C  d \theta_D^2
+ \sin^2 \theta_C \phi_C^2 +\cos^2  \theta_C \sin^2 \theta_D d \phi_D^2$ 
 and thereby with Eq. (\ref{laplacebeltrami} ), 
 the part of the Laplace operator which does not 
 commute with $ Tr  \lambda Q = 4 \lambda_C \lambda_D $ 
 is given by Eq. ( \ref{orth}), 
\begin{eqnarray} 
 \Delta^R_Q & = & \partial_{\lambda_C} 
 ( 1 - \lambda_C^2 ) \partial_{\lambda_C}
 + 2 \frac{1 - \lambda_C^2}{\lambda_C}   \partial_{\lambda_C}
\nonumber \\ 
& + & \frac{1}{\lambda_C^2}  \partial_{\lambda_D} 
 ( 1 - \lambda_D^2 ) \partial_{\lambda_D},
\end{eqnarray}
 where $\lambda_i = \cos \theta_i, i= C,D$.

For moderately strong 
spin- orbit scattering $1 / \tau_{SO} > \Delta_C$,
  in  the functional integral 
representation of the spectral determinants by 
  Grassman vectors the  spin degree of freedom is introduced,  
 $\alpha=2$  and   the matrix $C$ is, due to the time reversal 
 of the spinor,  substituted 
 by $i \sigma_2 \tau_1$\cite{elk}.
  The 
 spin-orbit scattering   reduces the Q matrix to unity in spin space.
 Thus, the matrix  C has effectively the form $\tau_1$.
   The condition $Q^T C = C Q$ leads therefore to a new symmetry class, 
 when the spin symmetry is broken
 but the 
 time reversal symmetry remains intact.  Then, $Q$ are $4 \times 4$-
 matrices on the orthogonal 
 symmetric space $ O(4)/(O(2) \times O(2) )$
 \cite{weg}.
 
 A matrix $Q$ with the above symmetries can be represented as,
\begin{equation}
Q= V^{-1} Q_0 V,
\end{equation}
with
\begin{equation}
 Q_c^0 = \left( \begin{array}{cc} \cos \hat{\theta} &  \sin \hat{\theta}
 
\\   \sin \hat{\theta}   & - \cos\hat{\theta}  \end{array} \right),
\end{equation}
where
\begin{equation}
 \hat{\theta} = \left(\begin{array}{cc}  \theta_1 & 
 \theta_2 \\  \theta_2 &  \theta_1
\end{array} \right),
\end{equation}
with
 $\theta_i \in [0,\pi], i=1,2$, 
and
\begin{equation}
V = \left(\begin{array}{cc} V_1 & 0
 \\ 0 & V_2
\end{array} \right),
\end{equation}
 where
\begin{equation}
V_i =  \exp ( i \chi_{i} \tau_3 ),
\end{equation}
 with 
 $\chi_i \in [0, 2 \pi], i=1,2$. 

 Thus, we find $ ds^2 = Tr Q^2/4 = \sum_{i=1,2} d\theta_i^2
 +  d {\bf \chi}^T \hat{g_{\chi}} {\bf \chi}$, 
where 
\begin{equation}
  \hat{g_{\chi}} = \left(\begin{array}{cc} \sin^2 \theta_1 + \sin^2 \theta_2
  & 
- \sin^2 \theta_1 + \sin^2 \theta_2
  \\ - \sin^2 \theta_1 + \sin^2 \theta_2  &  \sin^2 \theta_1 + \sin^2 \theta_2
\end{array} \right),
\end{equation}

 Thus,  the part of the Laplace operator which does not 
 commute with $ Tr  \lambda Q = 4 \lambda_1 \lambda_2 $ 
 is given by Eq. ( \ref{spo}), 
\begin{equation} 
\Delta^R_Q =\sum_{l=1,2} \partial_{\lambda_l} 
 ( 1 - \lambda_l^2 ) \partial_{\lambda_l},
\end{equation}
 where $\lambda_i = \cos \theta_i, i=1,2$.


\begin{references}
\bibitem{reviews}
P.A. Lee, T.V. Ramakrishnan, Rev. of Mod. Phys. {\bf 57}, 287 (1985);
{B. Kramer} and A.~MacKinnon, Rep. Prog.Phys. {\bf 56}, 1469 (1993);
 C. W. J. Beenakker, Rev. Mod. Phys. {\bf 69}, 731 (1997).
\bibitem{ef}
K. B. Efetov, 
{\it Supersymmetry in Disorder and Chaos} Cambridge University Press, 
Cambridge (1997). 
\bibitem{larkin} K. B. Efetov, A. I. Larkin, Zh. Eksp Teor.
 Fiz. {\bf 85}, 764(1983) ( Sov. Phys. JETP {\bf 58}, 444 ). 
\bibitem{khavin}
 M. E. Gershenson et al., 
Phys. Rev. Lett. {\bf 79}, 725(1997);
  Phys. Rev. {\bf B 58}, 8009 (1998).
\bibitem{altaronov} B. L. Altshuler, A. G. Aronov, Pis'ma
  Zh. Eksp. Teor. Fiz. {\bf 33}, 515 (1981)[JETP Lett. {\bf 33},499 (1981)]. 
\bibitem{dugaev}
 V. K. Dugaev, D.E.
Khmel'nitskii, Sov. Phys. JETP {\bf 59}, 1038 (1984).
\bibitem{aronov}
B. L. Altshuler,
A.G. Aronov in {\it Electron-electron interactions in disordered
  systems}, North Holland (1985).
\bibitem{beenakker} C. W. J. Beenakker, H. van Houten, Phys. Rev. {\bf B 37},
  6544 (1988).
\bibitem{haake} F. Haake, M. Kus, H.- J. Sommers, H. Schomerus, and
  K. Zychowski, J. Phys. A {\bf 29}, 3641(1996). 
\bibitem{us} S. Kettemann, D. Klakow, U. Smilamsky, J.
 Phys. A,3643(1997).
\bibitem{dyson} F. J. Dyson, J. Math. Phys. {\bf 3}, 1199 (1962).
\bibitem{efetov1982} K. B. Efetov, Zh. Eksp. Teor. Fiz. {\bf
    83}, 833 (1982)[Sov. Phys. JETP  {\bf 56},467 (1982)]; J. Phys. 
{\bf C 15}, L 909 (1982).  
\bibitem{altshk}
L. P. Gor'kov, O. N. Dorokhov, and F. V. Prigara, Zh. Eksp. Teor. Fiz. {\bf
    84}, 1440 (1983)[Sov. Phys. JETP  {\bf 57}, 838 (1983)].
 B. L. Alt'shuler, B. I. Shklobskii, Zh. Eksp. Teor. Fiz. {\bf
    91}, 127 (1986)[Sov. Phys. JETP  {\bf 64},127 (1986)].
\bibitem{isa} B. L. Altshuler, I. Kh. Zharekeshev, S. A. Kotochigova, and
  B. I. 
 Shklovskii,  Zh. Eksp. Teor. Fiz. {\bf
    94}, 343 (1988)[Sov. Phys. JETP  {\bf 67}, 625 (1988)];
 S. N. Evangelou, and E. N. Economou, Phys. Rev. Lett. {\bf 68}, 361 (1992);
 E. Hofstetter, and M. Schreiber, Phys. Rev. Lett. {\bf 73}, 3137 (1994);
 I. Kh.
  Zharekeshev, and  B. Kramer, Phys. Rev. Lett. {\bf 79}, 717(1997).
\bibitem{altfuchs} A. Altland, D. Fuchs, Phys. Rev. Lett. {\bf 74},
  4269(1995).
\bibitem{mirlin} A. D. Mirlin, Phys. Rep. 326, 259 (2000),
 Procy. of Intern. School of Physics " Enrico Fermi", Course CXLIII, IOS Press,
 Amsterdam ( 2000 ).
\bibitem{guhr}
T. Guhr, A. Müller-Groeling, and H.A. Weidenmüller, Phys. Rep. 299, 189 (1998).
\bibitem{prb} S. Kettemann,   Phys. Rev.  {\bf B 59  },  4799 (1999).
\bibitem{ab} E. Abrahams, P. W. Anderson, D. C. Licciardello, V. Ramakrishnan,
 Phys. Rev. Lett { \bf 42} 673(1979).
\bibitem{gorkov} L. P. Gor'kov, A. I. Larkin, D. E. Khmel'nitskii, Pis'ma
 Zh. Eksp. Teor. Fiz. {\bf 30 }, 248 (1979) ( JETP Lett. {\bf 30}, 228
 (1979)). 
\bibitem{khm} B. L. Altshuler, D. E. Khmelmitskii, A. I. Larkin, and
  P. A. Lee, 
 Phys. Rev. {\bf B 22}, 5142 (1980). 
\bibitem{elk} K. B. Efetov, A. I. Larkin, D. E. Khmel'nitskii,
 Zh. Eksp. Teor. Fiz. {\bf 79},
1120(1980) 
( Sov. Phys. JETP {\bf 52}, 568(1980) ). 
\bibitem{schmid} S. Chakravarty, A. Schmid, 
Physics Reports {\bf 140}, 193(1986).
\bibitem{phase2} B. L. Altshuler, A. G. Aronov, D.E. Khmelnitskii, J. Phys. 
C {\bf 15}, 7367 (1982).
\bibitem{phase1} M. E. Gershenson, Ann. Phys. {\bf 8},559 (1999). 
\bibitem{buettikerphase} M. Buettiker, cond-mat/0105519 (2001). 
\bibitem{hikamiln} S. Hikami, A. I. Larkin and Y. Nagaoka,
  Prog. Theor. Phys. {\bf 63}, 707 (1980).
\bibitem{weg} F. Wegner, Z. Physik { \bf B 36},l209(1979); 
Nucl. Phys. B {\bf 316}, 663(1989);
 S. Hikami, Prog. Theor. Phys. {\bf 64}, 1466 (1980).
\bibitem{prl} S. Kettemann and A. Tsvelik, Phys. Rev. Lett. {\bf 82},
3689 (1999).
\bibitem{prbr} S. Kettemann,   Phys. Rev. Rapid Commun.  {\bf B 62  }, R13282
  (2000).
\bibitem{lie} S. Helgason, {\it Differential Geometry and Symmetric
 Spaces} Academic Press, New York (1962).
\bibitem{zirnbauer} 
M. R. Zirnbauer Jy.  Math.  Phys. {\bf 37},4986 (1996). 
\bibitem{tbfm} M. Titov, P. W. Brouwer, A. Furusaki, C. Mudry,
 cond-mat/0011146.
 \bibitem{taras1} D. Taras- Semchuk, K. B. Efetov, Phys. Rev. Lett. 
 {\bf 85}, 1060 (2000); cond-mat/0010282 (2001). 
\bibitem{blanter} Ya. M. Blanter, A. D. Mirlin, B. A. Muzykantskii,
cond-mat/0011498 (2000).
 \bibitem{zirnbauer2} 
M. R. Zirnbauer
Phys. Rev. Lett. {\bf 69},1584 (1992),
A. D. Mirlin, A. Mullergroeling, M. R. Zirnbauer,
 Ann. Phys. ( New York) {\bf  236},  325 (1994);
P. W. Brouwer, K. Frahm, 
    Phys. Rev.   {\bf B 53} ,1490 (1996); B. Rejaei, Phys. Rev. {\bf B 53}, 
 R13235 (1996), B. Rejaei, Phys. Rev. {\bf B 53}, R13235 (1996).  
 \bibitem{mottdavis} N. F. Mott, E. A. Davis, 
{\it Electronic Processes in Non-crystalline Materials}, 
Clarendon Press, Oxford (1971).
\bibitem{andreev} A. V. Andreev and B. D. Simons
    Phys. Rev. Lett. 75, 2304-2307 (1995).
\bibitem{bouchaud} J. P. Bouchaud, J. Phys. 1 (France) {\bf 1}, 985 (1991).
\bibitem{lerner} I. V. Lerner, Y. Imry, Europhys. Lett. {\bf 29}, 49(1995).
\bibitem{leadbeater}  M. Leadbeater,
                               V. I. Falko, C.
                               J. Lambert,Phys. Rev.
                                           Lett. 81, 1274 (1998).
\bibitem{crossover}
J.-L. Pichard, M. Sanquer, K. Slevin, and P. Debray, Phys. Rev. Lett. 65, 1812
(1990); H. Schomerus and C.W. Beenakker, Phys. Rev. Lett. 84, 3927 (2000);
 M. Weiss, T. Kottos, T. Geisel, Phys. Rev. {\bf B 63}, R081306(2001).
\bibitem{kolesnikov} A.V. Kolesnikov and K.B. Efetov, Phys. Rev. Lett. 83, 3689 (1999); 
    A.V. Kolesnikov and K.B. Efetov, cond-mat/0005048 (to be published).
\bibitem{Miller} A. Miller and E. Abrahams, Phys. Rev. {\bf 120 }, 745 
(1960). 
\bibitem{Shkl1} B. I. Shklovskii and A. L. Efros, Electronic Properties of
Doped Semiconductors, Springer-Verlag, New York (1984).
\bibitem{Ambeg} V. Ambegaokar, B. I. Halperin and J. S. Langer, Phys. Rev.
{\bf B4  }, 2612 (1971).
\bibitem{Pollak} M. Pollak, J. Non-Crystal. Solids {\bf 11 }, 1 (1972).
\bibitem{Shkl2} B. I. Shklovskii and A. L. Efros, Zh. Eksp. Teor. Fiz.
{\bf 60  },  867 (1971) [Sov. Phys. JETP {\bf 33 }, 468 (1971)].
\bibitem{Mott} N. F. Mott, J. Non-Crystal. Solids {\bf 1 }, 1 (1968).
 \bibitem{Lee} P. A. Lee, Phys. Rev. Lett.{\bf 53 }, 2042 (1984);
R. A. Serota, R. K. Kalia and P. A. Lee, Phys. Rev. {\bf B 33 }, 8441 (1986).
\bibitem{Raikh1} M. E. Raikh and I. M. Ruzin, Zh. Eksp. Teor. Fiz. 
{\bf 95 }, 1113 (1989) [Sov. Phys. JETP {\bf 68 }, 642 (1989)].
\bibitem{Kurki} J. Kurkijarvi, Phys. Rev. {\bf B8 }, 922 (1973).
\bibitem{Raikh2} M. E. Raikh and A. L. Efros, Pis'ma Zh. Eksp. Teor. Fiz. 
{\bf 45 }, 225 (1987) [JETP Lett. {\bf 45 }, 280 (1987)]. 
\bibitem{Larkin} A. I. Larkin and D. E. Khmel'nitskii, Zh. Eksp. Teor. Fiz. 
{\bf 83 },  140 (1982) [Sov. Phys. JETP {\bf 56 }, 647 (1982)].
\bibitem{khavin2} Yu. B. Khavin, M. E. Gershenson, A. L. Bogdanov,
Phys. Rev. Lett. {\bf 81}, 1066 (1998).  
\bibitem{altland} A.
  Altland, S. Iida, K. B. Efetov, {\em J. Phys. A} {\bf 26} (1993)
  3545.
\bibitem{aleiner} I. L. Aleiner, A. I.  Larkin, Phys. Rev. {\bf B 54}, 
 14423 (1996). 
\bibitem{muzy} B. A. Muzykantskii, D. E. Khmelnitskii, JETP Letters
{\bf 62},76 (1995)].
\bibitem{aasa} A. V. Andreev,
O. Agam, B. D. Simons, and B. L. Altshuler, Phys. Rev. Lett. {\bf 76}, 
 3947 (1996). 
 Nucl. Phys. {\bf B 482}, 536 (1996). 
\bibitem{simons} B. D. Simons, O. Agam, A. V. Andreev, 
 J. Math. Phys. {\bf 38}, 1982 (1997). 
\bibitem{altland} A.
  Altland, S. Iida, K. B. Efetov, {\em J. Phys. A} {\bf 26} (1993)
  3545.
\end{references}
\end{document}